%% file: main.tex
\documentclass[12pt]{article}
\usepackage[utf8]{inputenc}
\usepackage[linkcolor=blue, citecolor=blue]{hyperref}       %
\include{math}  

\usepackage{mathrsfs}
\usepackage{bm}
\usepackage{multirow} 
\usepackage{longtable}
\usepackage{geometry}
\usepackage{graphicx}
\usepackage{listings}
\usepackage{enumerate}
\usepackage{amssymb}
\usepackage{url}
\usepackage[sort&compress]{natbib}
\usepackage{tabularx}
\usepackage{pdfpages}
\usepackage{emptypage}
\usepackage{float}
\usepackage{authblk}
\usepackage{makecell}
\usepackage{subcaption}
\usepackage{mathtools}
\hypersetup{
     colorlinks = true,
     linkcolor = blue,
     anchorcolor = blue,
     citecolor = blue,
     filecolor = blue,
     urlcolor = blue}

\usepackage{xcolor}
\usepackage[shortlabels]{enumitem}
\usepackage{changebar}

\definecolor{darkgreen}{RGB}{0,125,0}

\newcommand{\blind}{0}

\def\spacingset#1{\renewcommand{\baselinestretch}%
{#1}\small\normalsize} \spacingset{1}

\renewcommand{\th}{\theta}

\newcommand{\sig}{\sigma}
\newcommand{\om}{\omega}

\newcommand{\pis}{{\pi_\star}}
\newcommand{\phis}{\phi^\star}
\newcommand{\dhat}{\bar{\delta}_{95}}
\newcommand{\Ds}{D^\star}
\newcommand{\N}{N_{\alpha,\beta}}
\newcommand{\Qh}{\widehat{U}}
\newcommand{\Qs}{Q^\star}
\newcommand{\Uparam}{\widehat{U}}
\newcommand{\UQ}{U}

\newcommand{\Tmax}{T_\text{max}}

\newcommand{\Bin}{\mbox{Bin}}

\newcommand{\mm}{^{(m)}}

\newcommand{\ybar}{\overline{y}}

\renewcommand{\eqref}[1]{(\ref{#1})}

\newcommand\iid{\stackrel{\mathclap{\tiny\mbox{iid}}}{\sim}}

\begin{document}
\if0\blind
{
  \title{\bf A Comparative Tutorial of Bayesian Sequential Design and 
Reinforcement Learning}
    \author{Mauricio Tec$^a$%
    ,
    Yunshan Duan${^b}$, 
    and 
    Peter M\"uller$^{b}$ \\
    ${^a}$Department of Biostatistics, Harvard T.H. Chan School of Public Health \\
    $^b$Department of Statistics and Data Science, The University of Texas at Austin}
  \maketitle
} \fi

\if1\blind
{
  \bigskip
  \bigskip
  \bigskip
  \begin{center}
    {\LARGE\bf  A Comparative Tutorial of Bayesian Sequential Design and 
Reinforcement Learning}
\end{center}
  \medskip
} \fi

\newcommand{\dd}{\color{brown}}
\newcommand{\bb}{\color{black}}

\begin{abstract}
Reinforcement Learning (RL) is a computational approach to reward-driven learning in sequential decision problems. It implements the discovery of optimal actions by learning from an agent interacting with an environment rather than from supervised data. We contrast and compare RL with traditional sequential design, focusing on simulation-based Bayesian sequential design (BSD). Recently, there has been an increasing interest in RL techniques for healthcare applications. We introduce two related applications as motivating examples. In both applications, the sequential nature of the decisions is restricted to sequential stopping. Rather than a comprehensive survey, the focus of the discussion is on solutions using standard tools for these two relatively simple sequential stopping problems. Both problems are inspired by adaptive clinical trial design. We use examples to explain the terminology and mathematical background that underlie each framework and map one to the other. The implementations and results illustrate the many similarities between RL and BSD. The results motivate the discussion of the potential strengths and limitations of each approach.
\end{abstract}

\newpage
\spacingset{1.45} 

\section{Introduction}\label{sec:intro}

\begin{figure}[bth]
  \begin{center}
    \centering
    \begin{subfigure}[t]{.6\linewidth}\centering
    \centering
    \includegraphics[width=.9\textwidth]{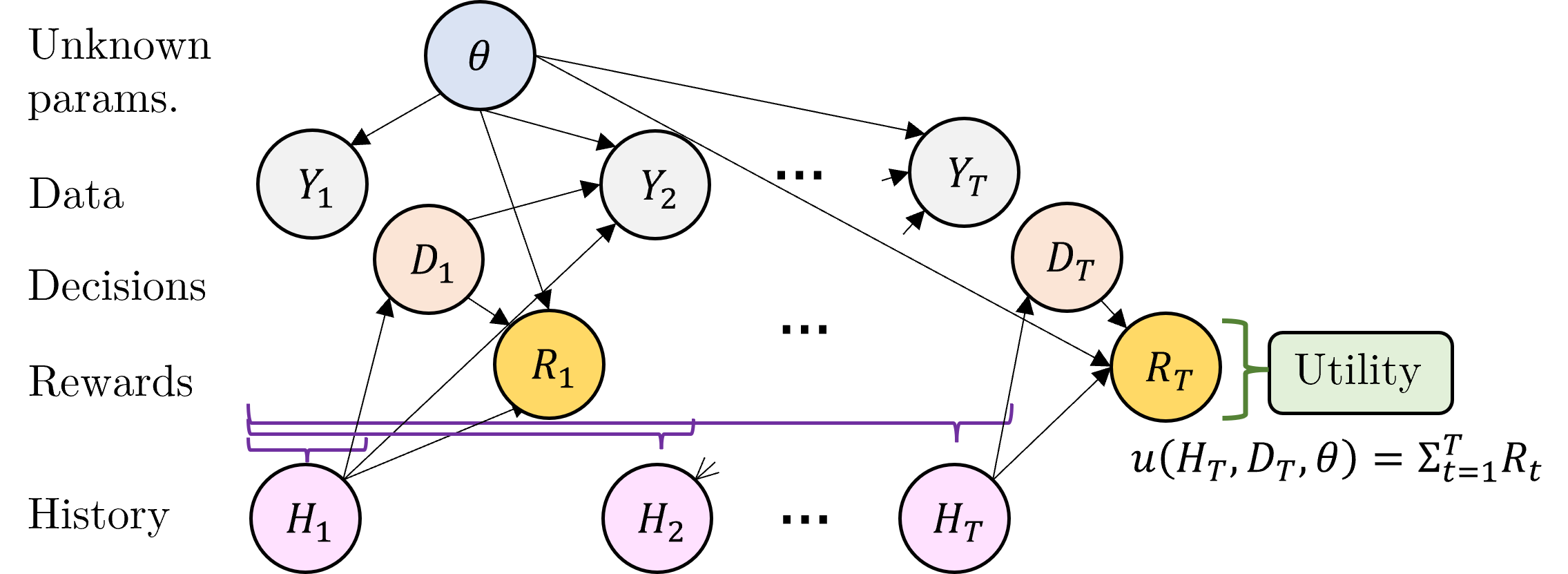}
    \caption{Generic SDP.}  
    \label{fig:diagrams:general}
    \end{subfigure}
    \begin{subfigure}[t]{.4\linewidth}\centering
    \centering
    \includegraphics[width=.9\textwidth]{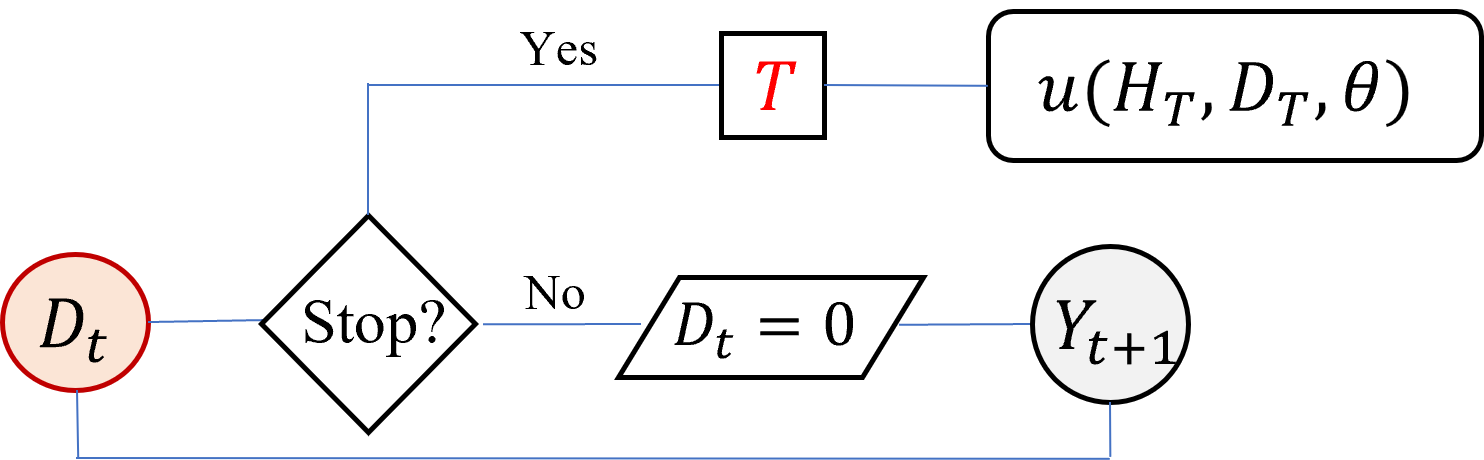}
    \caption{Sequential stopping.}
    \label{fig:diagrams:stopping}
    \end{subfigure}
    \caption{Panel (a) shows the directed graph of a general SDP
      problem.
      In the figure, $H_t = \{Y_1, D_1, Y_2, \ldots, D_{t-1}, Y_t\}$ denotes
      the history or information set at decision time $t$. There is an
      implicit arrow from every element of $H_t$ onto $D_{t}$,
      summarily implied by the arrow $H_{t}\to D_{t}$. 
      In RL the
      utility function is usually written as a sum 
      $G=\sum_{t=1}^T R_t$ of immediate rewards $R_t(H_t,D_t)$ (although it is usually called the \emph{return} instead of the utility).
      Panel (b) shows a flow chart for the special case of
      sequential stopping problems, with $D_t=0$ indicating
      continuation. 
      The node \fbox{\footnotesize${\color{red}T}$} represents
      stopping the trial ($t=T$ and $D_T\neq 0$).
    }
    \label{fig:diagrams:spd}
  \end{center}
\end{figure}

Sequential design problems (SDP) involve a sequence of
decisions $D_t$ with data $Y_t$ observed at every time step
$t=1,\hdots, T$ \citep{degr:70, berger2013statistical}. The goal is to
find a decision rule $(Y_1,\hdots, Y_t) \mapsto D_t$ that maximizes
the expected value of a utility function.
The utility function encodes an agent's
preferences as a function of hypothetical future data
and assumed truth. 
The decision rule is prescribed before observing
future data beyond $Y_t$ --
as the term ``design" emphasizes -- assuming a probabilistic model with
unknown parameters $\theta$ that generate the future data. For
example, $\theta$ can be the true effect of a drug.
Figure \ref{fig:diagrams:spd}a summarizes the setup of a general
SDP.
As motivating examples in the upcoming discussion we will use
two examples of clinical trial design, which naturally give rise to
SDPs \citep{Rossell:13b,
  BerryHo88, ChristenNakamura03}.
Both examples are about sequential
stopping, that is, the sequential decision is
to determine \textit{when} and \textit{how} to end the study.
Figure \ref{fig:diagrams:spd}b shows the setup of sequential stopping
problems.

Using these examples,
we compare two families of simulation-based methods for solving SDPs
with applications in sequential stopping.  The first is
simulation-based Bayesian Sequential Design (BSD)
\citep{muller2007simulation}, which is based on Bayesian decision
theory \citep{berger2013statistical}.  
The other approach is Reinforcement Learning (RL), a
paradigm based on the interaction between an agent and an environment
that results in potential rewards (or costs) for each decision.
RL has recently been proposed as a method to implement SDPs
focusing on recent advances in deep learning
\citep{shen2021bayesian}, 
 outside the context of clinical studies.
Earlier application of RL related to clinical study design
can be found in the dynamic treatment regimes literature
\citep{zhao2009reinforcement, murphy2003optimal,
murphy2007methodological}.
There is a longstanding 
literature on such problems in statistics; see the review by
\citet[chapter 15]{ParmigianiLurdes:09} and the references
therein. Implementing RL and BSD in these two motivating example problems (using
standard algorithms) will illustrate many similarities between
the two frameworks, while highlighting the potential strengths and
limitations of each paradigm\footnote{The implementation code is freely available at \url{https://github.com/mauriciogtec/bsd-and-rl}.}.

A note about the scope of the upcoming discussion. It is meant to
highlight the similarities and differences of algorithms in BSD and RL,
and we provide a (partial) mapping of notations. There is no intent to
provide (yet another) review of BSD or RL and its many variations.
This article focuses on a few variations that best contrast
the two traditions,
\cbstart
and we try to highlight the relative advantages of each method.
In short, BSD is better equipped to deal with additional structure,
that is, to include more details in the inference model. For example, dealing
with delayed responses in a clinical study, one
might want to include a model to use early responses to predict such
delayed outcomes. Or one might want to borrow strength across multiple 
related problems. Doing so would also be possible in RL, but it requires to leave the framework of Markov decision processes underlying many RL algorithms \citep{gaon2020reinforcement}, and discussed below. On the other hand, RL allows computation-efficient implementations,
and is routinely used for much larger scale problems than BSD.
The availability of computation-efficient implementations is 
critical in applications where sequential designs need to be evaluated
for summaries under (hypothetical) repeated experimentation.
This is the case, for example, in clinical trial design when frequentist
error rates and power need to be reported. The evaluation requires
Monte Carlo experiments with massive repeat simulations, making
computation-efficient implementation important.
\cbend

\paragraph*{BSD:}
BSD is a model-based paradigm for SDPs based on a sampling model
$p(Y_t \mid H_{t-1}, D_{t-1}, \theta)$ for the observed data and a prior
$p(\theta)$ reflecting the agent's uncertainty about the unknown
parameters. 

To compare alternative decisions $D_t$, the agent uses an optimality
criterion that is formalized as
a utility function $u(Y_1,\hdots, Y_T, D_1,\hdots,
D_T, \theta)$ which quantifies the agent's preferences under hypothetical data, decisions and truth. 
It will be
convenient to write the utility as $u(H_T, D_T=d, \theta)$, where
$H_t$ denotes the history $H_t:=(Y_1,D_1, \ldots, D_{t-1}, Y_t)$ 
at decision time $t$.
Rational decision makers should act as if they were to maximize such
utility in expectation conditioning on already observed data, and
marginalizing with respect to any (still) unknown
quantities like future data or parameters \citep{degr:70}. 

To develop a solution strategy, we start at time $T$ (final
horizon or stopping time). 
Denote by $\UQ(H_T,d) = E\{u(H_T, D_T=d, \theta) \mid H_T\}$ the
expected utility at the stopping time $T$.
Then, the rational agent would  select $\Ds_T(H_T)=\argmax_d
\UQ(H_T, d)$ at the stopping time. For earlier time steps, rational
decisions derive from expectations over future data,
 with later optimal decisions plugged in, 
\begin{equation}\label{eq:DTs}
\UQ(H_t, d) = E\{u(H_{t},D_t=d, Y_{t+1}(D_t=d), \Ds_{t+1}(H_{t+1}(D_t=d)), \hdots, \Ds_T(H_T(D_t=d)), \theta \mid H_t \}
\end{equation}
which determine the optimal decision (``Bayes rule'') as
\begin{equation}\label{eq:BR}
\Ds_{t}(H_t)  =  \textstyle{\argmax_d\,} \UQ(H_{t}, d). 
\end{equation}
Here, we use a \emph{potential outcomes} notation $H_{t+1}(D_t=d)$ to emphasize that future history depends on the action $D_t=d$
\citep{robins1997causal}, and similarly for other quantities.
In the notation for expected utility $U(\cdot)$, the lack of
an argument indicates marginalization (with respect to
future data and parameters) and optimization (with respect to future
actions), respectively. For example, in \eqref{eq:DTs} expected
utility  conditions on $H_t$, but marginalizes w.r.t. future data
$Y_{t+k}$ and substitutes optimal future decisions $\Ds_{t+k}$. 

Some readers, especially those familiar with
RL, might wonder why $D_t=d$ does not appear on the right-hand side of
the conditional expectation (as in RL's state-action value
functions).
 This is because in the BSD framework actions are deterministic.
There is no good reason why a rational decision maker would
randomize \citep{berger2013statistical}.
\cbstart However, note that in clinical studies randomization is usually
included and desired, but for other reasons -- not to achieve an optimal
decision, but to facilitate attribution of differences in outcomes to
the treatment selection.  \cbend

\paragraph{RL} RL addresses a wider variety of sequential problems
than BSD, provided one can formulate them as an agent interacting with
an environment yielding rewards $R_t$ at every time step. Environments
can be based on simulations. For example, popular successful RL applications with
simulation-based environments include Atari video-games
\citep{mnih2013playing}, complex board games like chess and Go \citep{silver2018general}, robotic tasks
\citep{tunyasuvunakool2020dm_control} and autonomous driving
\citep{wurman2022outracing, sallab2017deep}.  

Just as in BSD, the interactive setup for RL (transition and rewards) can be defined by a 
sampling model and a prior over $\theta$. The interaction replicates
the decision process, shown in Figure \ref{fig:diagrams:spd}.
Each draw $\theta \sim p(\theta)$
constitutes a new instance or \emph{episode} of the environment. The
RL agent seeks to maximize the expected sum of rewards $G=\sum_{t=1}^T
R_t$, known as the \emph{return}, over an episode. The return $G$ is
the analogue of the utility function.

Decision rules are called \emph{policies} in RL. A (stochastic)
policy maps an observed history to a distribution over actions $D_t
\sim \pi(\cdot \mid H_t)$. The optimal policy $\pis$ satisfies
$\pis=\argmax_\pi E\{G \mid \pi\}$. As mentioned before,
the notion of stochastic policies is not natural in BSD with its
focus on decisions that a rational agent would take.
In fact, under some regularity conditions it can be shown 
that also the optimal RL policy is deterministic
\citep{puterman2014markov}. So why stochastic
policies?
 Stochastic policies in RL serve to implement exploration. In BSD
it is assumed that if exploration were called for, it would be
recognized by the optimal decision rule. While in theory this is the
case, in practice additional reinforcement of exploration is
reasonable.
Also, as we shall see later, the use of stochastic policies
facilitates the search for optimal policies by allowing the use of
the stochastic gradient theorem. 
 
Another close similarity of BSD and RL occurs in the
definition of expected utility and the state-action value function in RL.
The \emph{state-action value function} 
is $Q^\pi(H_t, d)= E\{\sum_{k=t}^T R_k \mid H_t, D_t=d, \pi\}$. The
value function of the optimal policy $Q^{\pis}(H_t,d)$ plays the same
role as expected utility $\UQ(H_t, d)$  when optimal decisions are
substituted for future decisions $D_s$, $s>t$. An important difference
is the stochastic nature of $\pi$ in the state-value function, versus
the deterministic decisions $D_t$ in BSD. 

From this brief introduction one can already notice many
correspondences between the objects in RL and BSD. Table
\ref{tab:notation} shows a partial mapping between their respective
terminologies.  Not all are perfect equivalences. Sometimes common use
in BSD and RL involves different levels of marginalization and/or
substituting optimal values.  Some of the analogies in the table will
be developed in the remainder of the paper.

\newcommand{\both}[1]{\multicolumn{2}{c}{#1}}
\begin{table}[p]
  \small
  \caption{A brief
    comparison of key quantities in BSD and RL.
    Variations without time subindex $_t$ refer to time-invariant
    versions. Using states $S_t$, in many instances an argument $H_t$
    can be replaced by $S_t$, as in $D_t(S_t)$,
    We use $Y=(Y_1,\hdots, Y_T)$ etc. to refer to lists over $t=1,\ldots,T$.}
  \label{tab:notation}
  \begin{center}
    \spacingset{1.0}
    {\renewcommand{\arraystretch}{1.2} 
      \begin{tabular}{l|p{5.8cm}|p{5.8cm}}
        & BSD & RL \\
        \hline
        $Y_t$ & \both{data observed at time $t$ }\\
        $H_t=(Y_1,D_1 \ldots D_{t-1}, Y_t)$ & \both{history (information
                                              set) at decision time $t$}\\
        $D_t=D_t(H_t)$ & \both{action (decision) at time $t$}\\
        $S_t=S_t(H_t)$ & summary (sufficient) statistic & state \\
        $D_{\phi,t}$ & (deterministic) action at time $t$ & n/a $^{(a)}$ \\
        & indexed by parameter   $\phi$ (policy) &  \\
        $\pi(D_t \mid H_t)$ & n/a (no randomization) & (random)  policy \\
        $\pi_\phi$ & n/a &  policy indexed by $\phi$ \\
        optimal decision/policy & Bayes rule $\Ds(H_t)$ \eqref{eq:BR}
              & optimal policy $\pis(H_t)$ \\
        $\th$ & \both{unknown parameter }\\
        & (usually) required & optional \\
        $R_t$ & n/a $^{(b)}$ & (immediate) reward at time $t$\\ [4pt]
        optimality criterion &
              utility $u(Y, D, \theta)=u(H_t,D_t, \theta)$ &
              total return $G=\sum_{k=1}^T R_t$ or \\
        & & remaining return $G_t=\sum_{k=t}^T R_k$\\[4pt]
        state-action value & n/a (deterministic $\Ds$) &
              $Q^{\pi}(H_t,D_t)= E_\pi\{G_t \mid H_t, D_t\}$ \\[4pt]
        state value  & n/a (deterministic $\Ds$)  &  $V^{\pi}(H_t)=E_\pi\{G_t \mid H_t, \pi\}$ \\ [4pt]
        value under optimal  & $\UQ(H_t, D_t)$ & $Q^{\pis}(H_t, D_t)$  \\
        ~~~ future actions & & \\
        optimal value & $\UQ(H_t, \Ds(H_t))$ &  $V^{\pis}(H_t)$ \\ [4pt]
        $J(\phi)$ & \both{expectation under policy/decision indexed by $\phi$}\\
        &  $= E\{u(Y, D_{\phi}, \theta)\}$
              & $= E\{G \mid \pi_\phi\}$  \\ [4pt]
        Bellman equation/ & $U(H_t, D_t)=$ & $Q^{\pi}(H_t, D_t)=$ \\
        backward induction &~~~ $E\{U(H_{t+1}(D_{t})) 
                             \mid H_t\}$ & ~~ $E\{R_t + V^{\pi}(H_{t+1})\mid H_t, D_t,\pi\}$
      \end{tabular}}
  \end{center}
  $^{(a)}$  deterministic policies $\pi_\phi(H_t)$ are not discussed in this review, but see
  \citet{silver2014deterministic}.\\
  $^{(b)}$ additive decomposition 
  as $u(H_T,D_t,\th) = \sum_t R(H_t, D_t,\th)$ is possible, but not
  usually made explicit.
\end{table}

\section{Two examples of optimal stopping in clinical trials}
\label{sec:examples}

The two stylized examples introduced here mimic
sequential stopping in a clinical trial.
The agent is an investigator who is planning and overseeing the trial. The data $Y_t$ are clinical outcomes recorded for each patient.
In both cases $D_t$ refers to a stopping decision after $t$ (cohorts of)
patients. Under continuation ($D_t=0$), the agent incurs an additional cost
$c_t$ for recruiting the next cohort of patients. Under stopping ($D_t
\ne 0$), on the other hand, the agent incurs a cost if a wrong (precise meaning
to be specified) recommendation is made.
At each time the agent has to choose between continuing the study -- to
learn more -- versus stopping and realizing a reward for a good final
recommendation (about the treatment).
 Throughout we use the notions of cost (or loss) and utility
interchangeably, treating loss as negative utility. 

\paragraph*{Example 1: A binary hypothesis}\label{sec:example1}

Consider the decision problem of choosing between $H_1: \theta =
\theta_1$ and $H_2: \theta = \theta_2$. For instance, $\theta$ could
represent the probability of a clinical response for an experimental
therapy. Assume a binary outcome $Y_t$ with a Bernoulli sampling model
$p(Y_t=1 \mid \theta) = \theta$ and a discrete two-point prior
$p(\theta = \theta_1) = p(\theta = \theta_2) = \frac{1}{2}$.

\begin{figure}[tbhp]
  \centering
  \begin{subfigure}[t]{0.32\linewidth}
    \centering
  \includegraphics[width=\linewidth]{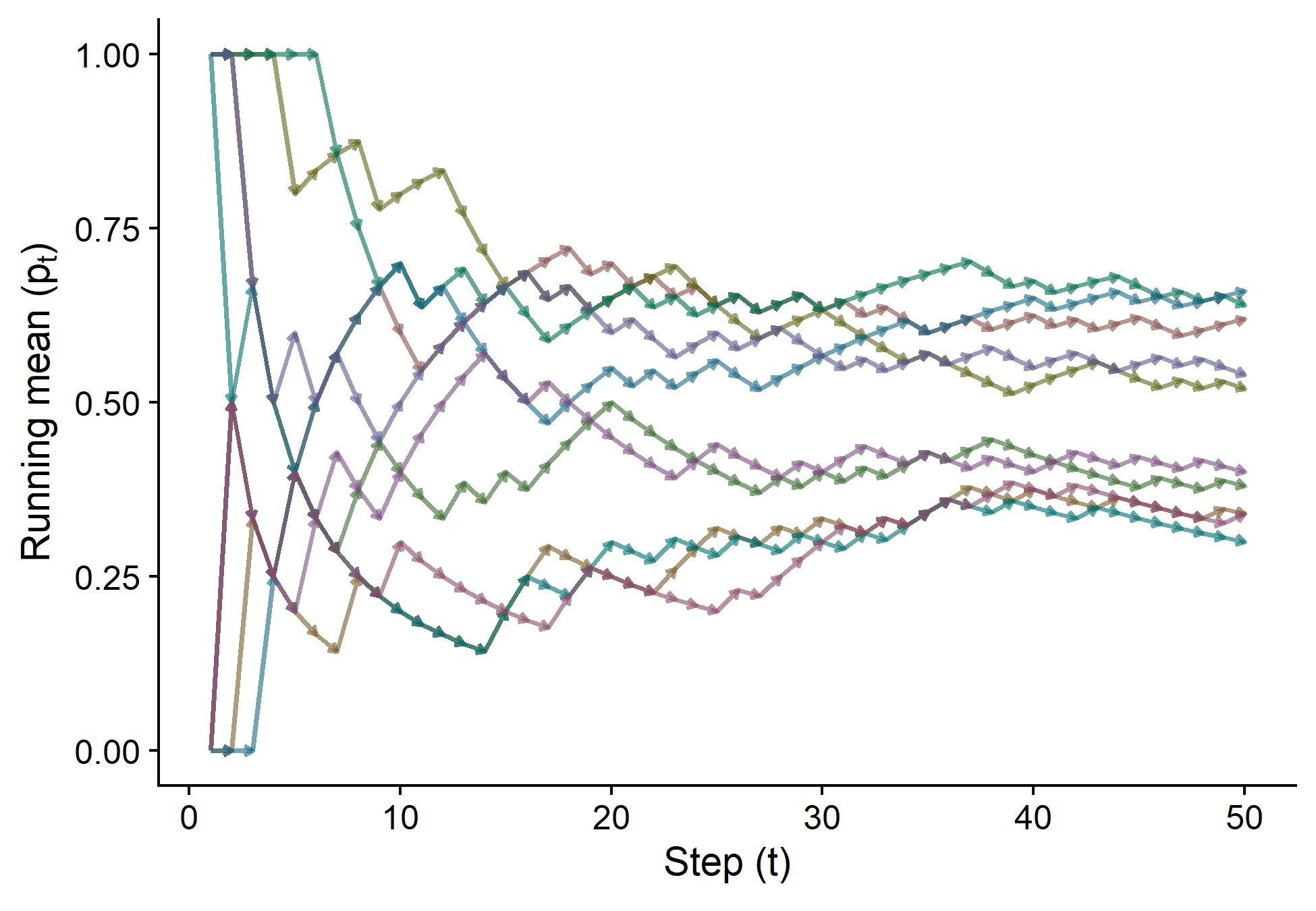}
  \caption{Example 1: trajectories}
  \label{fig:data:1}
  \end{subfigure}\hfill
  \begin{subfigure}[t]{0.32\linewidth}
    \centering
  \includegraphics[width=\linewidth]{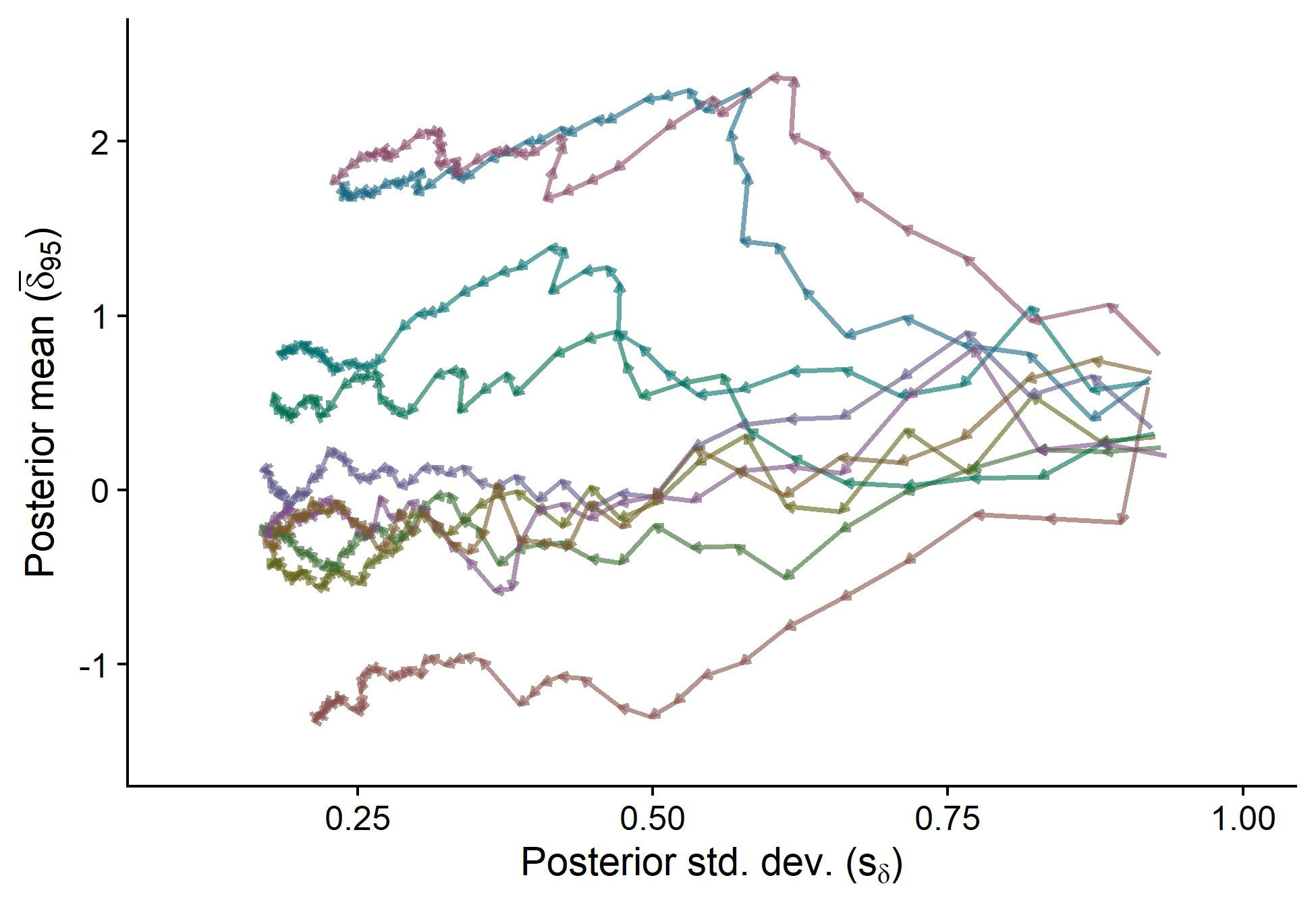}
  \caption{Example 2: trajectories}
  \label{fig:data:2}
  \end{subfigure}
  \begin{subfigure}[t]{0.35\linewidth}
    \centering
  \includegraphics[width=0.85\linewidth]{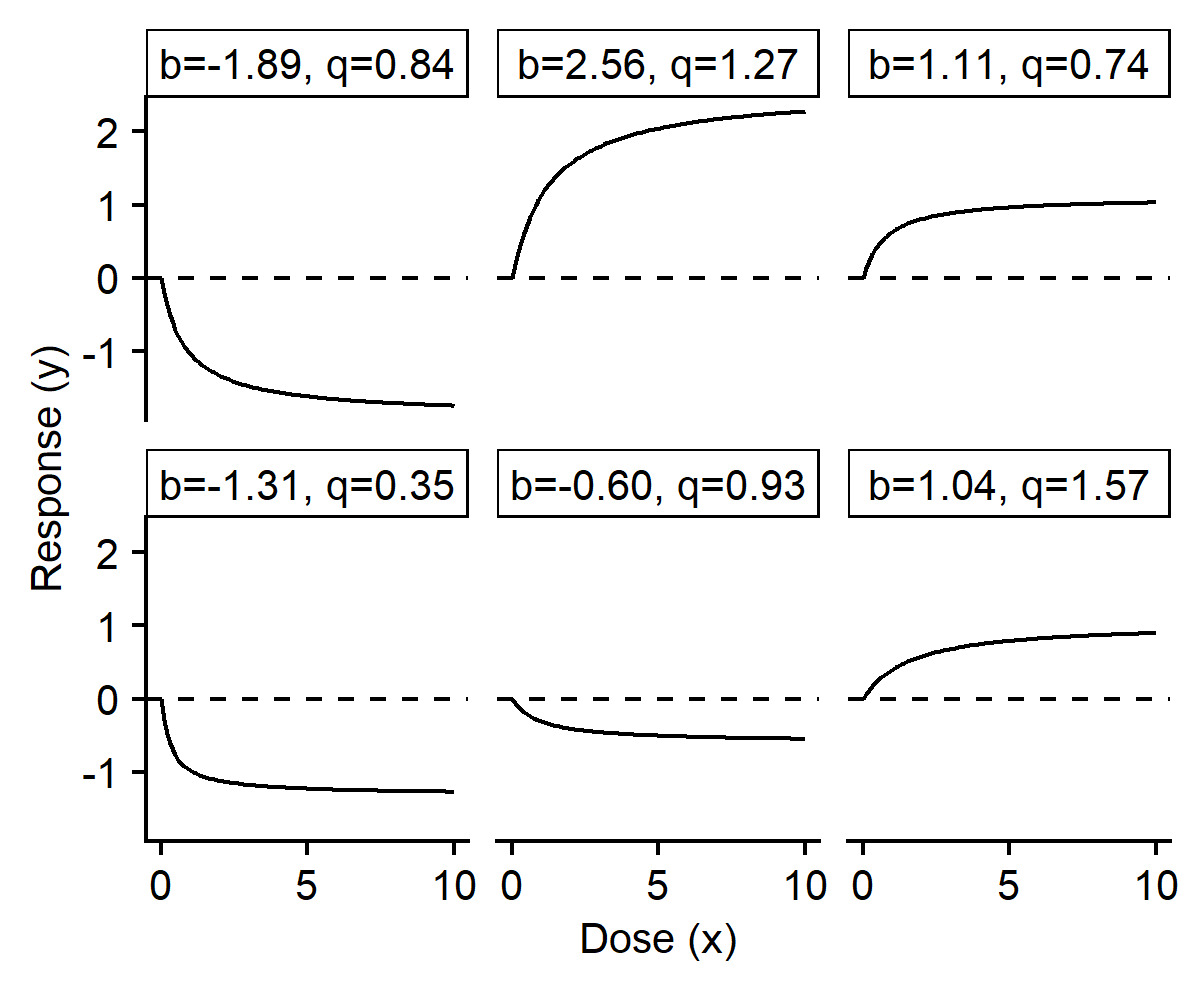}
  \caption{Example 2: dose-response models}
  \label{fig:data:3}
  \end{subfigure}\\
  \caption{Data from the two sequential stopping examples. (a) and (b)
    are forward simulations assuming no stopping;  (c) shows the
    implied dose-response curves for different random draws of the prior.}  
  \label{fig:data}
\end{figure}

The possible decisions at any time are $D_t \in \{0,1,2\}$.
Here $D_t=0$ indicates continuation, 
$D_t = 1$ indicates to terminate the trial and report $H_1
(\theta=\theta_1)$, and 
$D_t = 2$ means terminate and report $H_2 (\theta=\theta_2)$. The utility function includes a (fixed) sampling cost $c$ for each cohort and a final cost $K>0$ for reporting the wrong hypothesis. The utility function is
\begin{equation}\label{eq:utility1}
u(H_T,D_T, \theta) = - cT - K\, \mathbb{I}(\theta\neq \theta_{D_T}).
\end{equation}
The relevant history $H_t$ can be represented using the summary
statistic $S_t=(t,\sum_{k\le t} Y_k / t)$, since this statistic is
\emph{sufficient} for the posterior of $\theta$. The implementations
of simulation-based BSD and RL use this summary statistic. The problem
parameters are fixed as $c=1$ and $K=100$. Example trajectories of
$S_t$ assuming no stopping are shown in Figure \ref{fig:data:1}.

\paragraph*{Example 2: A dose-finding study}\label{sec:example2}
This example is a stylized version of the ASTIN trial
\citep{grieve2005astin}. The trial aims to find the optimal dose for a
drug from a set of candidate doses $\sX=\{\rx_0,\hdots,\rx_G\}$ where
$\rx_0=0$ stands for  placebo. At each time $t$, a dose $X_t\in\sX$
is assigned to the next patient (cohort), and an efficacy outcome
$Y_t$ is observed. The aim is to learn about the dose-response curve
$f(X_t) = \E(Y_t \mid X_t)$, and more specifically, to find the dose
$\rx_{95}$ (the ED95) that achieves 95\% of the maximum possible
improvement over placebo. Let $\delta_g=f(\rx_g) - f(\rx_0)$ be the
advantage over the placebo at dose $\rx_g$.
We set up a nonlinear regression
$Y_t = f(X_t \mid \theta) + \epsilon_t$ with $\epsilon_t \sim N(0,
\sigma^2)$ using a dose-response function 
\begin{equation}\label{eq:dose-response}
  f(\rx \mid \theta)  = a + b \frac{\rx^r}{(q^r + \rx^r)}
\end{equation}
with $\theta =(a, b, q, r)$ and a prior $p(\theta) = N(\theta_0, diag(\lambda_0^2))$. 
In the PK/PD (pharmacokinetics/pharmacodynamics) literature model
\eqref{eq:dose-response} is known as the $E_{max}$ model
\citep{meibohm1997basic}.

Similar to Example 1,
$D_t \in \{0,1,2\}$ with 
$D_t=0$ indicating continuation,
$D_t=1$ indicating stopping the trial and recommending no further
investigation of the experimental therapy, and
$D_t=2$ indicating stopping and recommending for a follow-up
trial set up as a pivotal trial to test the null hypothesis
$H_0:$ $\delta_{95}=0$.
If continuing the trial, the next assigned dose is
$X_{t+1}=\min\{X_t + \xi, \hat{x}_{95,t}\}$ where $\hat{x}_{95,t}$ is
the latest estimate of the ED95,
 and $\xi$ is a maximum allowable dose escalation between
cohorts. 
If requiring a pivotal trial, $\N$ patients are assigned to the dose
$\hat{x}_{95,T}$, with $\N$ computed from the observed data to achieve
a desired power $1-\beta$ at a certain alternative $H_1:
\theta=\theta_1$ for test of size $\alpha$. Details are given in Appendix A.

The utility function includes a patient recruitment cost of $c$ and a
prize $K>0$ if the null hypothesis $H_0(\delta_{95}=0)$ is rejected in
the pivotal trial (meaning the agent found evidence of an effective
drug). Denote $\Delta_R = \Pr(\mbox{reject $H_0$ in the 2nd trial}
\mid H_T)$.  At the stopping time $T = \min_t\{D_t \ne 0\}$,  
utility is calculated as
\begin{equation}\label{eq:utility2}
  u(H_T, D_T, \theta) =
  \begin{cases} -c T & \mbox{if } D_T=1\\
    -cT + \left\{-c \N(H_T) + K\Delta_R(H_T)\right\} & \mbox{if } D_T=2
  \end{cases}
\end{equation}
In our implementation we  fix $a=0,r=1, \sigma=1$ in \eqref{eq:dose-response} and 
$c=1,K=100$ in \eqref{eq:utility2}, 
and $\xi=1$, leaving the unknown parameters $\theta=(b, q)$, including
the maximum effect $b$ and the location of the $\rx_{50}$. The prior
of $\theta$ has mean $\theta_0=(1/2,1)$ with variances
$\lambda_0=(1,1)$ and we add the constraint $q\geq 0.1$.

The summary statistic is $S_t=(\dhat, s_\delta)$ where $\dhat$ and
$s_{\delta}$ are posterior mean and standard deviation of
$\delta_{95}$.
Figure \ref{fig:data:2} shows examples of trajectories
of these summaries  until some maximum time horizon $T$ (i.e., 
assuming no stopping). The trajectories are created by sampling $\theta$ from
the prior, assigning doses as described, and sampling responses
using \eqref{eq:dose-response}.
Figure \ref{fig:data:3} shows examples of the implied
dose-response curve $f(\rx\mid\theta)$ for different prior
draws. Notice that the chosen summaries do not capture the full
posterior. The statistic $S_t$ is not a sufficient statistic for
the posterior.  However, as shown in Appendix A, $\N$ and
$\Delta_R$, and therefore the utility, depend on the data only through
$S_t$.

\section{Simulation-based Bayesian sequential design}
\label{sec:solutions-bsd}

From the expected utility definition in \eqref{eq:DTs}, one immediately deduces that
\begin{equation}\label{eq:bellman-bsd}
\UQ(H_t, D_t=d) = E\{\UQ(H_{t+1}(D_t=d), \Ds(H_{t+1}(D_t=d))) \mid H_t\},
\end{equation}
 with the expectation being with respect to future data and $\th$,
and substituting optimal choices for future decisions, $s>t$.
In words,  for a rational agent taking optimal actions,
the expected utility given history $H_t$ must be the same as the
expected utility in the next step. Thus, one can (theoretically)
deduce $\Ds_t$ from knowing the best actions for all possible future
histories  by implementing {\em backward induction} starting with
$T$.  
Equation \eqref{eq:bellman-bsd} is known as the \emph{Bellman equation}
\citep{bellman1966dynamic}. 

Enumerating all histories is computationally intractable in most
realistic scenarios, rendering backward induction usually
infeasible, except in some special setups \citep{BerryHo88,
ChristenNakamura03}.
Simulation-based BSD comes to help: instead of enumerating all
possible histories, we compute approximations using some simulated
trajectories. The version presented here follows
\cite{muller2007simulation}.  Similar schemes are developed in
\cite{BrockwellKadane03}, \cite{KadaneVlachos02} and
\cite{carlin1998approaches}.
We use two strategies: first,
we represent history $H_t$ through a (low-dimensional) summary
statistic $S_t$, as 
already hinted in Section \ref{sec:examples} when we
proposed the posterior moments of the ED95 response for Example 2. The
second -- and closely related -- strategy is to restrict $D_t$
to depend on $H_t$ only indirectly through $S_t$.
Two instances of this approach are discussed below and used to
solve Examples 1 and 2.

\subsection{Constrained backward induction}
\label{sec:CBI}
Constrained backward induction is an algorithm consisting of
three simple steps.
The first step is \emph{forward simulation}. Our implementation here uses the assumption that the sequential nature is limited to sequential stopping, so trajectories can be generated assuming no stopping independently from decisions. Throughout we use $D_t=0$ to denote
continuation.  Other actions, $D_t \ne 0$, indicate stopping the
study and choice of a terminal decision.
The second step 
is constrained \emph{backward induction}, which implements
\eqref{eq:bellman-bsd} using decisions restricted to depend on the history
$H_t$ indirectly only through $S_t$.
The third step simply keeps track of the best action and iterates until convergence. We
first briefly explain these steps and then provide an illustration of
their application in Example 1 and additional implementation
considerations. 

\begin{description}
\item [ Step 1.] {\em Forward simulation:} Simulate many trajectories,
  say $M$, until some maximum number of steps $\Tmax$ (e.g. cohorts in
  a trial). To do this, each $m=1,\hdots, M$ corresponds to a
  different prior draw $\th\mm \iid p(\th\mm)$ and samples
  $Y\mm_t\iid p(Y\mm_t \mid \th\mm)$, $t=1,\hdots, \Tmax$.
  For each $m$ and $t$, we evaluate and record the summary
  statistic $S\mm_t$ discretized over a grid.  

\item [Step 2.] {\em Backward induction:}
For each possible decision $d$ and each grid value $S=j$, the
algorithm approximates $\Qh(S,d)\approx U(S, d)$ using the forward
simulation and Bellman equation as follows. Denote with $A_j=\{(m, t_m)
\mid S\mm_{t_m}=j\}$ the set of forward simulations that
fall within grid cell $j$. Then, 
\begin{equation}
  \Qh(S=j, d) =
  \begin{cases}
  \frac{1}{|A_{j}|} \sum_{(m, t_m) \in A_{j}}
  \Qh(S\mm_{t_m+1},\Ds(S\mm_{t_m+1})) & d=0\\
  \frac{1}{|A_{j}|} \sum_{(m, t_m) \in A_{j}}
  u(S_{t_m}\mm, D_{t_m}=d, \th\mm) & d \ne 0.
  \end{cases}
\label{eq:Qh}  
\end{equation}
The evaluation under $d=0$ requires the optimal actions
$\Ds_{t+1}(S_{t_m+1})$. We use an initial guess (see below), which is then iteratively
updated (see next).

\item [Step 3.] {\em Iteration:} Update the table
  $\Ds(S) = \argmax_d \, \Qh(S,d)$ after step 2.
\end{description}
\cbstart Repeat steps 2 and 3 until updating in step 3 requires no (or
below a minimum number of) changes of the tabulated $D^\star(S)$. \cbend

\begin{figure}[!tbhp]
  \centering
  \begin{subfigure}[t]{0.7\linewidth}
    \centering
    \includegraphics[width=\linewidth]{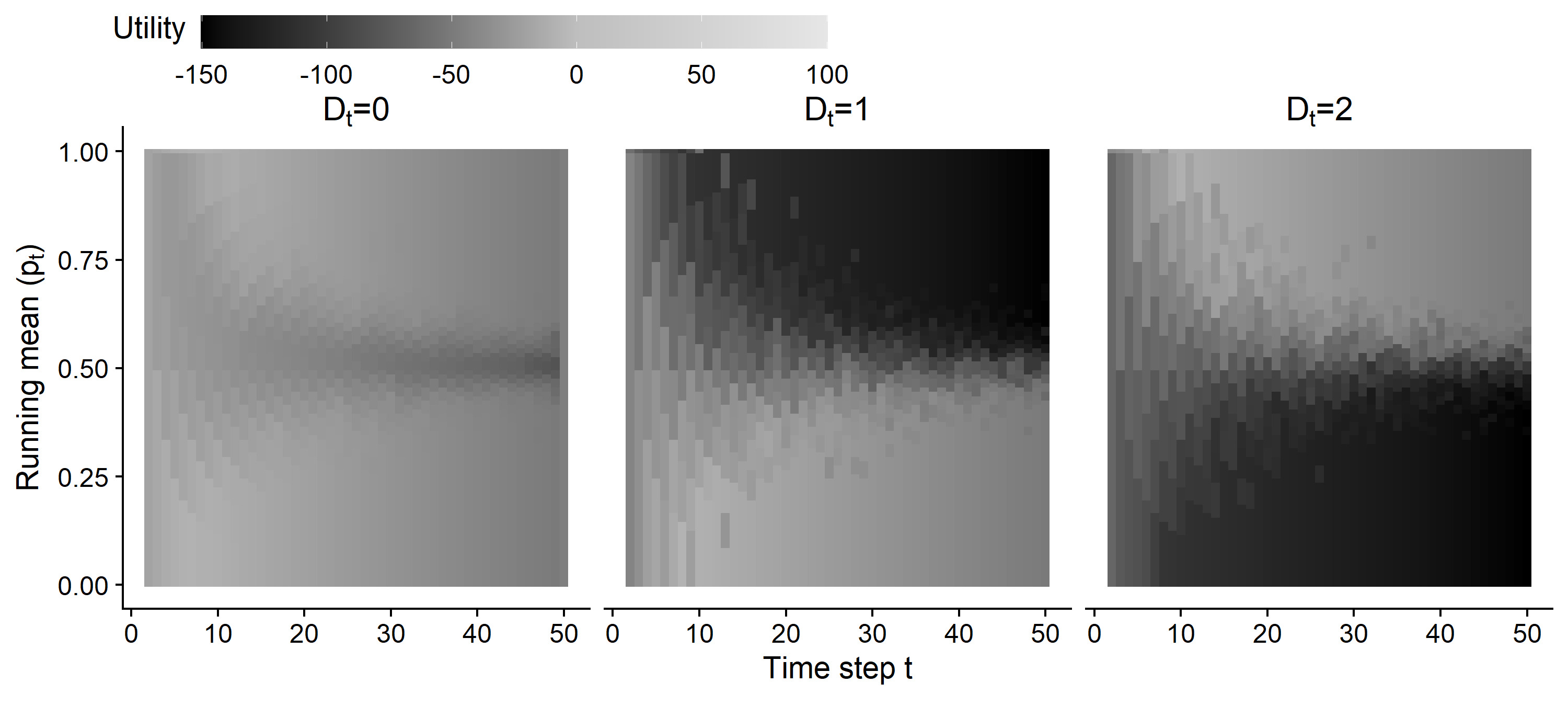}
    \caption{BSD: expected utility estimates under
      constrained backward induction.} 
    \label{fig:Eg1-value:bsd}
  \end{subfigure}\hfill%
  \begin{subfigure}[t]{0.3\linewidth}
    \centering
    \includegraphics[width=\linewidth]{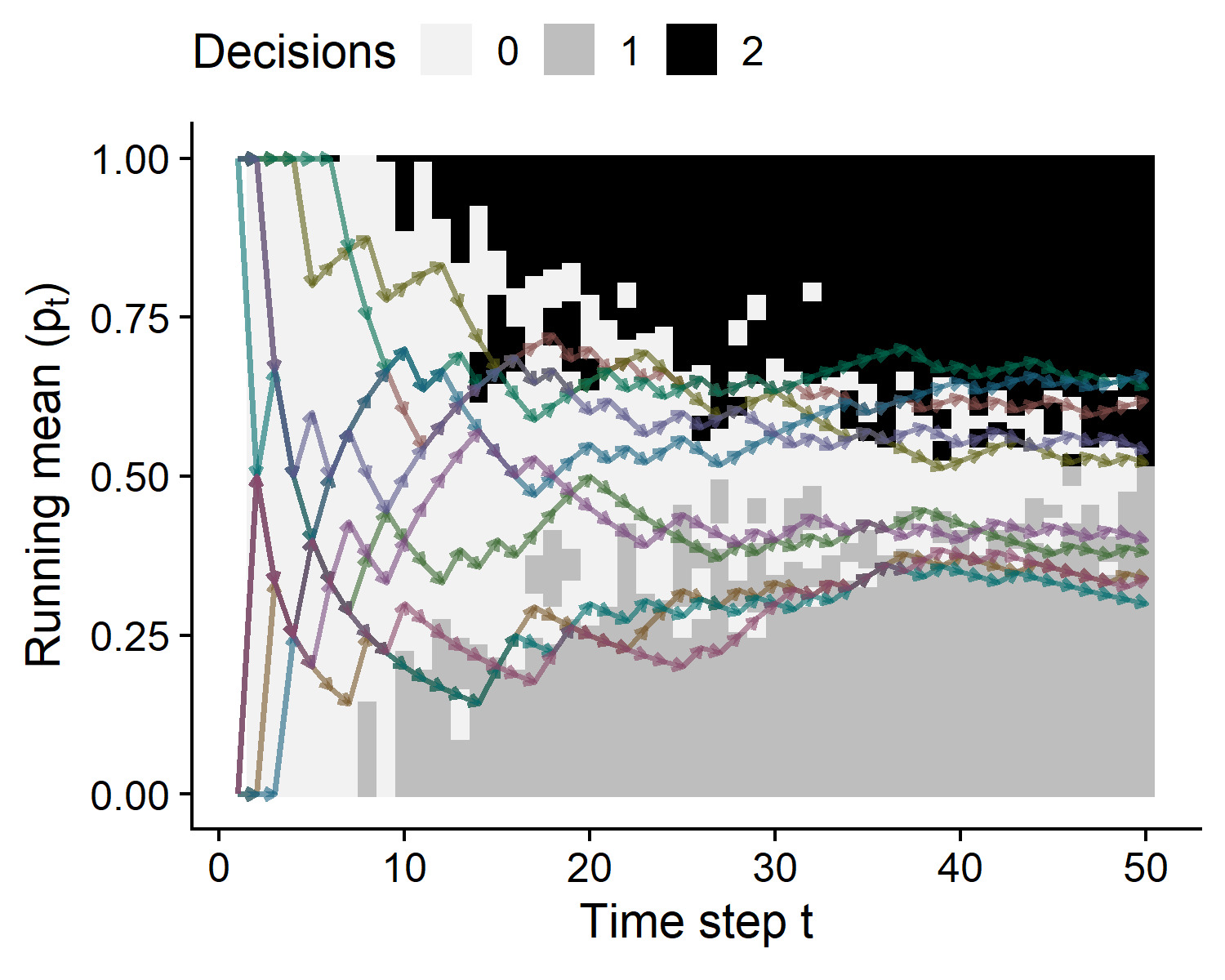}
    \caption{BSD: optimal actions.}
    \label{fig:Eg1-value:bsd-bounds}
  \end{subfigure}\\
  \begin{subfigure}[t]{0.7\linewidth}
    \centering
  \includegraphics[width=\linewidth]{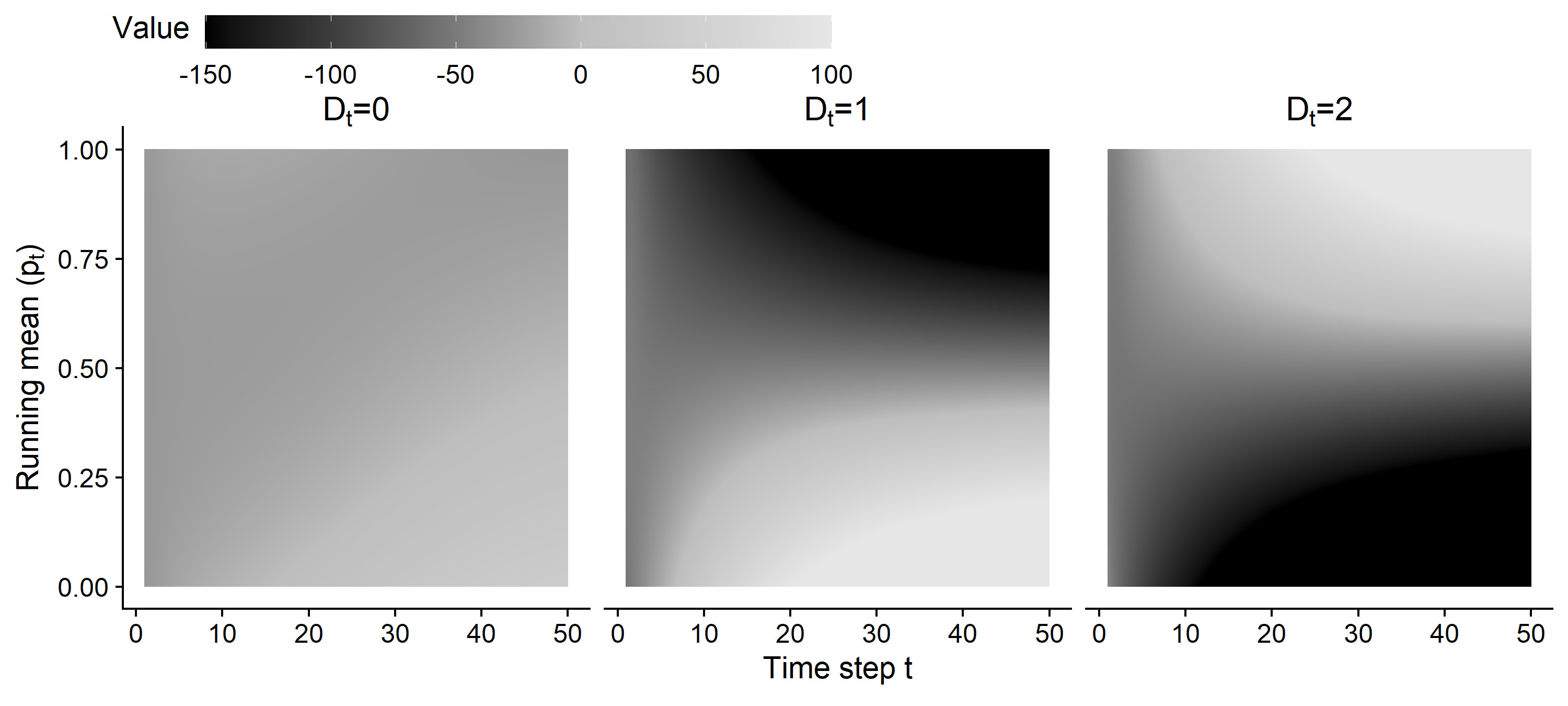}
  \caption{RL: state-action value estimates with Q-learning}
  \label{fig:Eg1-value:rl}
  \end{subfigure}\hfill
  \begin{subfigure}[t]{0.3\linewidth}
    \centering
  \includegraphics[width=\linewidth]{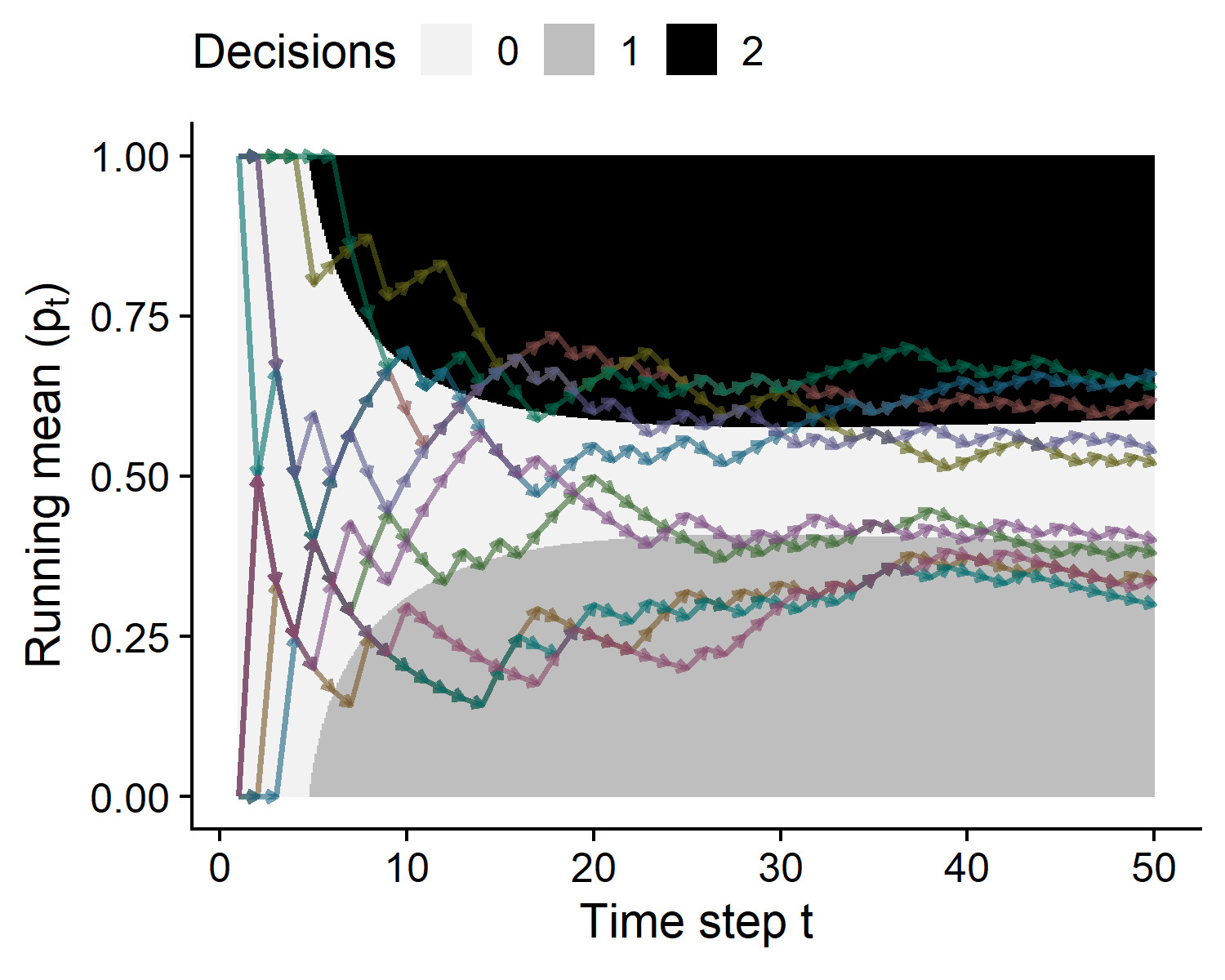}
  \caption{RL: best actions.}
  \label{fig:Eg1-value:rl-bounds}
  \end{subfigure}\\
\begin{subfigure}[t]{.33\linewidth}
\raggedright
\includegraphics[width=0.99\linewidth]{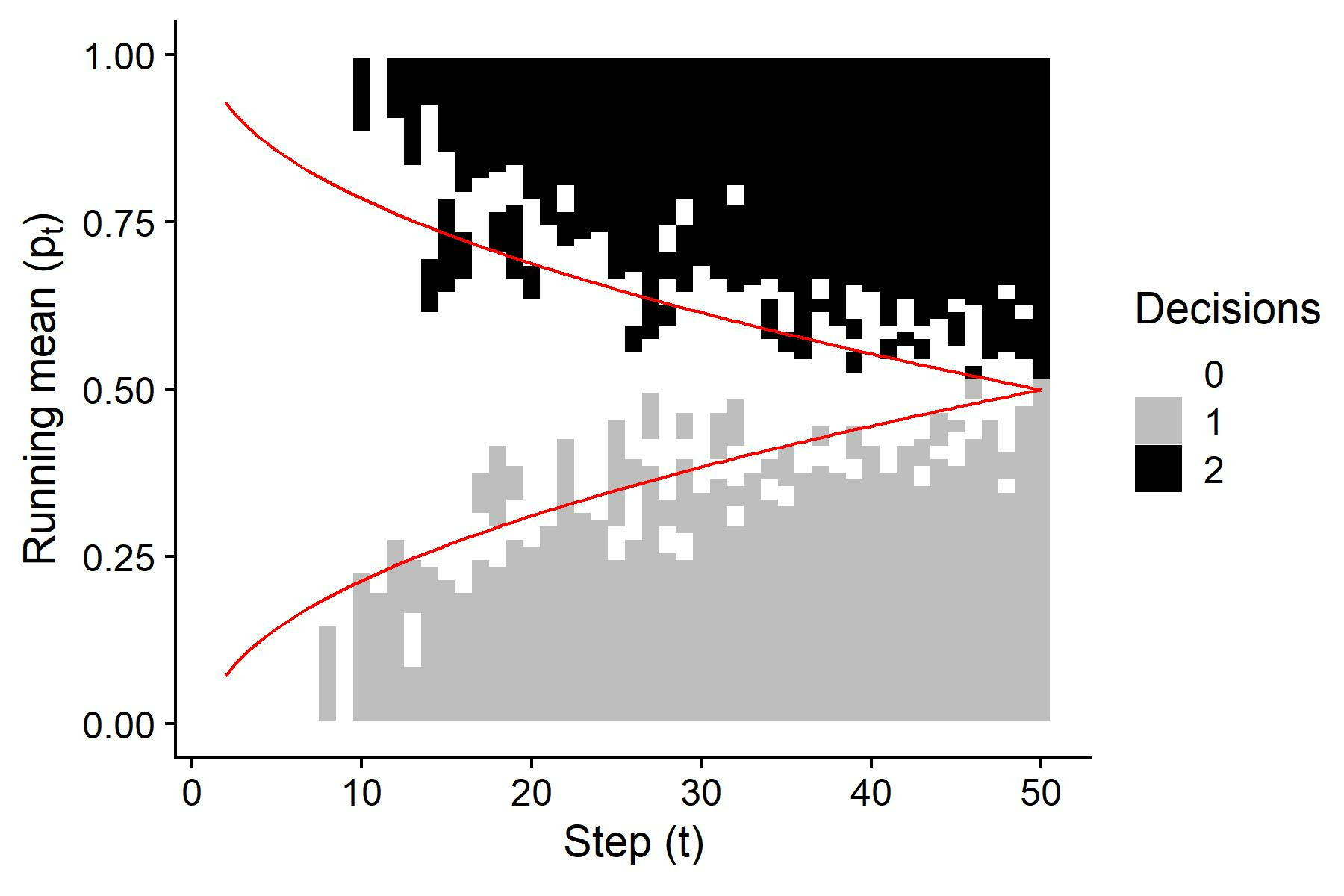}
\vspace*{-0.8cm}
\caption{BSD: parametric boundary.}
\label{fig:Eg1:parametric-bounds}
\end{subfigure}\hfill%
\begin{subfigure}[t]{.33\linewidth}
\raggedleft
\includegraphics[width=0.99\linewidth]{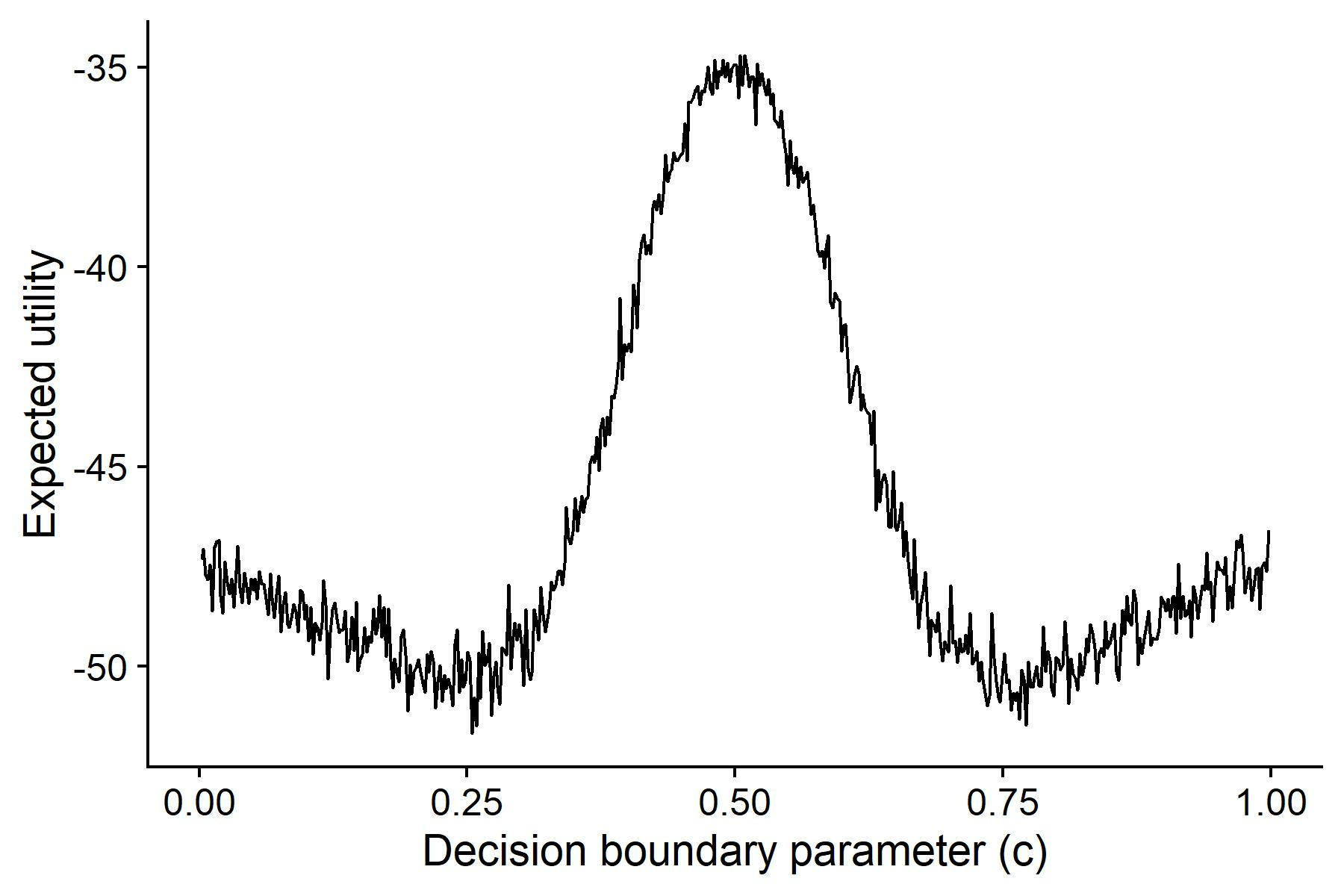}
\vspace*{-0.8cm}
\caption{BSD: utility by parameter.}
\label{fig:Eg1:parametric-values}
\end{subfigure}\hfill%
\begin{subfigure}[t]{.33\linewidth}
\raggedleft
\includegraphics[width=0.99\linewidth]{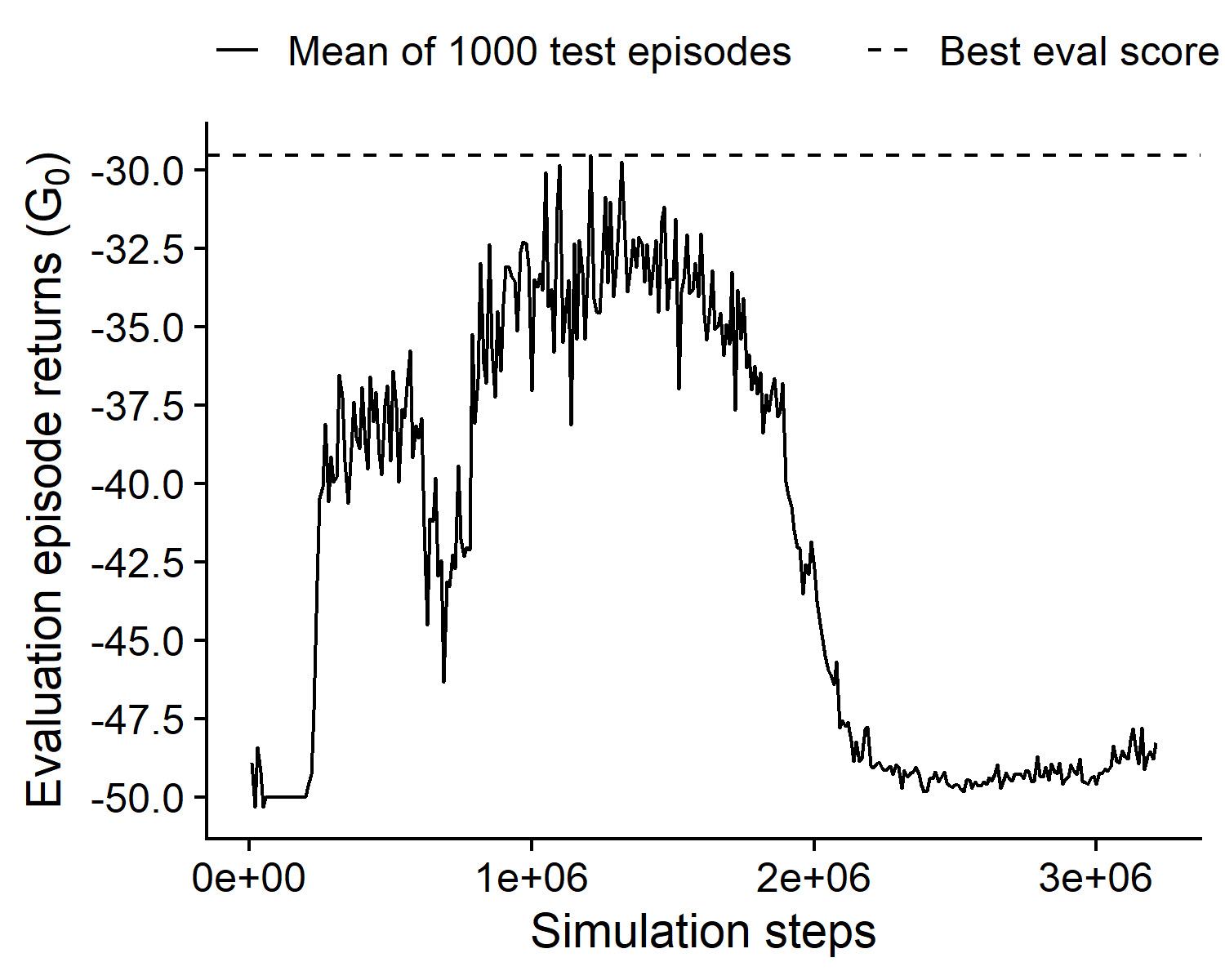}
\vspace*{-0.8cm}
\caption{RL: performance over training.}
\label{fig:Eg1:RL:training}
\end{subfigure}
\caption{Example 1. Comparison of decision boundaries and fitted value functions/utilities.} 
\label{fig:Eg1}
\end{figure}

Figure \ref{fig:Eg1-value:bsd} shows the estimated utility function
$\Qh(S, d)$  in Example 1 using $M=1000$, $\Tmax=50$ and 100 grid
values for the running average $p_t=\sum_{k=1}^t Y_t/t$ in $S=(t,
p_t)$.
Optimal actions $\Ds_t(S)$ are shown in Figure
\ref{fig:Eg1-value:bsd-bounds}. The numerical uncertainty due to the
Monte Carlo evaluation of the expectations is visible. If desired, one
could reduce it by appropriate smoothing
\citep{macdonald2015gpfit}. One can verify, however, that the
estimates are a close approximation to the analytic solution
which is available in this case
\citep{muller2007simulation}. 

 We explain Step 2 by example. Consider Figure
\ref{fig:Eg1-value:bsd-bounds} and assume, for example, that we need
the posterior expected utility of $S=(t,p_t)=(20,0.25)$. In this
stylized representation, only three simulations,
$A=\{m_1, m_2, m_3\}$ 
pass through this grid cell. 
In this case, $t_m=t=20$ for the three trajectories since $t$ is
part of the summary statistic. We evaluate $\Qh(S,d=1)$ and
$\Qh(S,d=2)$ as averages
$\frac1{3} \sum_{i \in A}u(S_{20}\mm,d,\th\mm)$.  
For $\Qh(S,d=0)$, we first determine the grid cells in the next
period for each of the three trajectories. 
$S\mm_{t_m+1}=(21,p\mm_{21})$. We then look up the optimal
decisions $\Ds(S_{21},p\mm_{21})$ (using in this case $t_m+1=21$) and average
$\frac13 \sum \Qh(21,p\mm_{21},\Ds(21,p\mm_{21}))$, as in \eqref{eq:Qh}.

Constrained backward induction
requires iterative updates of $\Ds(S)$ and their values.
The procedure starts with arbitrary initial values for $\Ds(S)$,
recorded on a grid over $S$. For example, a possible initialization is
$\Ds(S) = \max_{d  \ne 0} \Qh(S,d)$, maximizing over all actions that
do not involve continuation. With such initial values, $\Qh(S,d)$ can
be evaluated over the entire grid.
Then, for updating the optimal actions 
$\Ds(S)$ 
one should best start from grid values that are associated
with the time horizon $T$, or at least high $t$. This is particularly
easy when $t$ is an explicit part of $S_t$, as in Example 1 with
$S_t=(t,p_t)$. Another typical example arises in Example 2 with
$S_t=(\mu_t,\sig_t)$, the mean and standard deviation of some quantity of interest. For large $t$ we expect small $\sig_t$,
making it advisable to start updating in each iteration with the grid
cells corresponding to smallest $\sig_t$. We iterate until no more
(or few) changes happen. 

 The algorithm can be understood as an implementation of
\eqref{eq:bellman-bsd}. 

Consider $f(S, d)$ as an arbitrary function over pairs
$(S,d)$ and the function operator $\gP f$ defined as $(\gP f)(S,d) =
\max_{d'}\E[f(S',d') | S]$ where $S'$ is the summary statistic
resulting from sampling one more data point $Y$ from the unknown
$\theta$ and recompute $S'$ from $S$.
Then Bellman equation
\eqref{eq:bellman-bsd} can be written as $U=\gP U$. In other words,
expected utility  under the optimal decision $\Ds$ 
is a fixed point of the operator $\gP$.
 Constrained backward induction attempts to find an approximate
solution to the fixed-point equation. 
The same principle motivates the Q-learning algorithm in RL
\citep{watkins1992q} (see Section \ref{sec:rl}). Backward induction is also closely related to the \emph{value iteration} algorithm for Markov decision processes  \citep{sutton2018reinforcement}, which relies on exact knowledge of the state transition function.

\subsection{Sequential design with decision boundaries}

Inspection of Figure \ref{fig:Eg1-value:bsd-bounds} suggests an attractive
alternative algorithm. Notice the decision boundaries
on $S=(t,p_t)$ that trace a
funnel with an upper boundary $\om_2(t)$ separating $\Ds=2$ from
$\Ds=0$, and a lower boundary $\om_1(t)$ separating  $\Ds=0$ versus
$\Ds=1$.

Recognizing such boundaries suggests an alternative approach based on
searching for optimal boundaries in a suitable family $\{\om_{\phi,1},
\om_{\phi,2} \mid \phi \in \Phi\}$. 

This approach  turns the sequential decision
problem of finding optimal $\Ds(S)$ into a non-sequential problem of
finding an optimal $\phis \in \Phi$. This method is used, for example, in
\cite{Rossell:13b}.

In Example 1,  we could use 
$$
\om_{1}(t) = \frac{\phi\sqrt{t-1}}{\sqrt{T-1}},
\quad\quad
\om_{2}(t) = 1 - \frac{(1-\phi)\sqrt{t-1}}{\sqrt{T-1}}
$$
using a single tuning parameter $\phi \in (0,1)$.
Both functions are linear in $\sqrt{t-1}$, and mimic the funnel shape
seen in Figure \ref{fig:Eg1-value:bsd-bounds}. The decision rules
implied by these boundaries is
\begin{equation*}
  D_{\phi}(S_t) = 
  \begin{cases}
    1 & \mbox{ if } p_t < \om_1(t)\\ 
    2 & \mbox{ if } p_t > \om_2(t)\\ 
    0 & \mbox{ otherwise. }
  \end{cases}
\end{equation*}
Here, the additional subscript $_\phi$ in $D_{\phi}(\cdot)$
indicates that the decision follows the rule implied by decision
boundaries $\om_j(t; \phi)$. Note that $\om_1(1) = 1$ and
$\om_2(1) = 0$, ensuring continuation at $t=1$. 

For a given $\phi$, the forward simulations are used to evaluate
expected utilities under the policy $D_{\phi}$.  Let $T\mm_\phi =
\min\{t:\; D_{\phi}(S\mm_t \ne 0\}$ denote the stopping time under
$D_{\phi}$.
Then the expected utility under policy $D_\phi$ is
\begin{equation}
  U(\phi) =E\{u(S_{T_\phi},D_{\phi}(S_{T_\phi}),\theta)\}
\end{equation}
 where the expectation is with respect to data $Y_T$ and $\th$.
It is approximated as
an average over all Monte Carlo simulations, stopping each simulation at
$T\mm_\phi$, as determined by the parametric decision boundaries,

\begin{equation}
   \Uparam(\phi) = (1/M)\sum_{m=1}^M
   u(S\mm_{T^{m}_\phi},D_{\phi}(S\mm_{T^{m}_\phi}),\theta\mm).
  \label{eq:Qhphi}
\end{equation}
Optimizing $\Qh(\phi)$ w.r.t. $\phi$ we find the optimal decision
boundaries $\phis = \arg\max \Qh(\phi)$. 
As long as the nature of the sequential decision is restricted to
sequential stopping, the same set of Monte Carlo simulations can be used to evaluate all $\phi$, using different truncation to evaluate $\Qh(\phi)$. 
In general, a separate set of forward simulations for each
$\phi$,  or other simplifying assumptions  might be required.

Figure \ref{fig:Eg1:parametric-bounds} shows the decision boundaries
for the best parameter estimated at $\phis=0.503$ in Example 1. The
estimated values for $\Qh(\phi)$ are in Figure
\ref{fig:Eg1:parametric-values}.
In Figure \ref{fig:Eg1:parametric-bounds},
the boundaries using constrained backward induction are
overlaid in the image for comparison.
The decision boundaries trace the optimal decisions under the backward
induction well.
The differences in expected utility close to the decision boundary are
likely very small, leaving minor variations in the decision boundary
negligible.

The same approach is applied to the (slightly more complex)  Example
2.

Recall the form of the summary
statistics $S=(s_\delta, \bar{\delta})$, the posterior standard
deviation and mean of the ED95 effect. We use the boundaries 
$$
\omega_1(S) = -b_1 s_\delta + c,
\quad\quad
\omega_2(S) = b_2 s_\delta + c,
$$
parameterized by $\phi = (b_1,b_2,c)$. The implied decision rules are 
\begin{equation*}
D_\phi(S) = 
    \begin{cases}
    1 & \mbox{if } \bar{\delta} < \omega_1(s_\delta) \\
    2 & \mbox{if } \bar{\delta} > \omega_2(s_\delta) \\
    0 & \text{otherwise}.
    \end{cases}
\end{equation*}

\begin{figure}[!tbhp]
  \centering
  \begin{subfigure}[t]{.48\linewidth}
  \raggedright
  \includegraphics[width=0.99\linewidth]{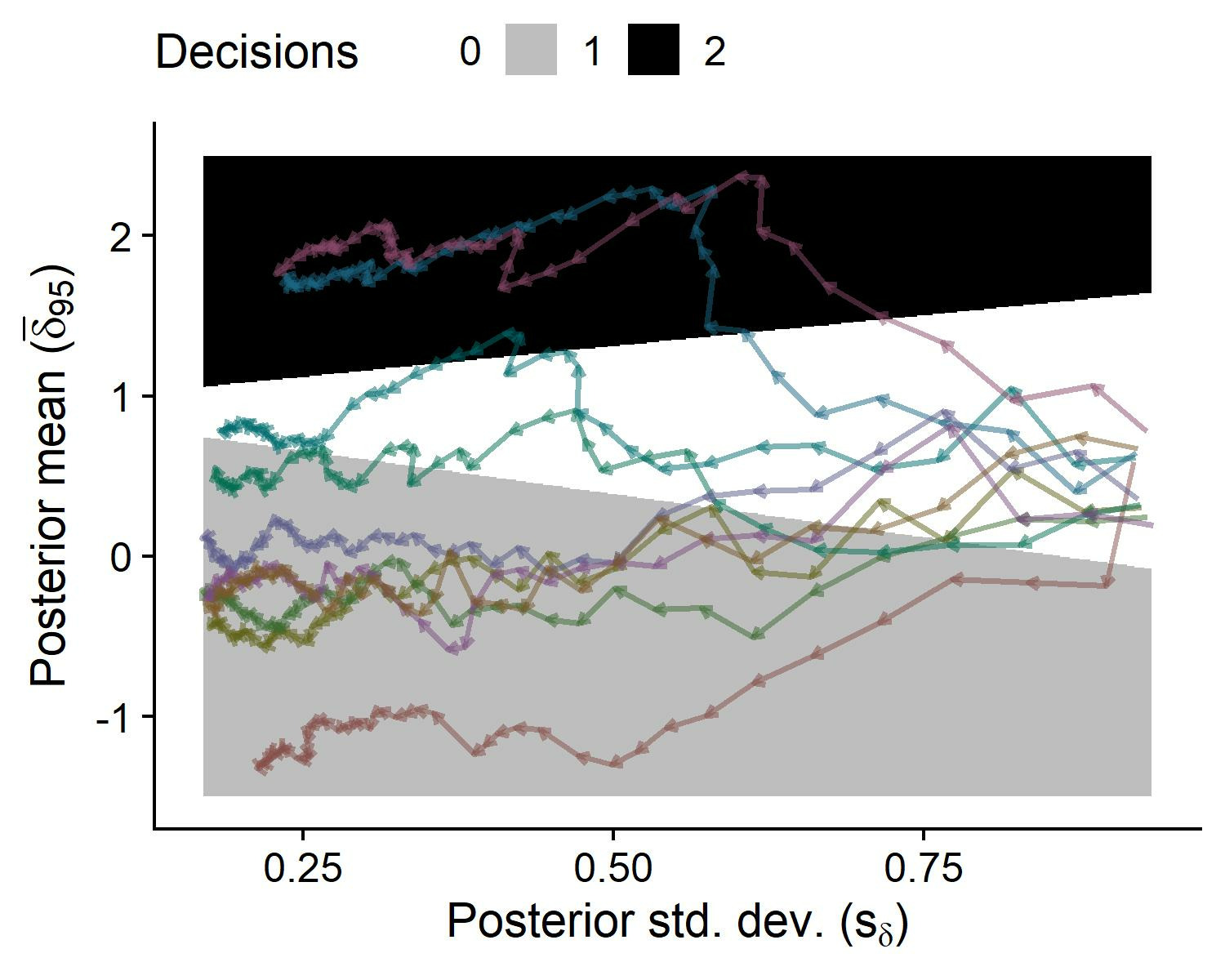}
  \vspace*{-0.8cm}
  \caption{BSD: parametric boundaries.}
  \label{fig:Eg2_bound:bsd}
  \end{subfigure}\hfill%
  \begin{subfigure}[t]{.48\linewidth}
  \raggedleft
  \includegraphics[width=0.99\linewidth]{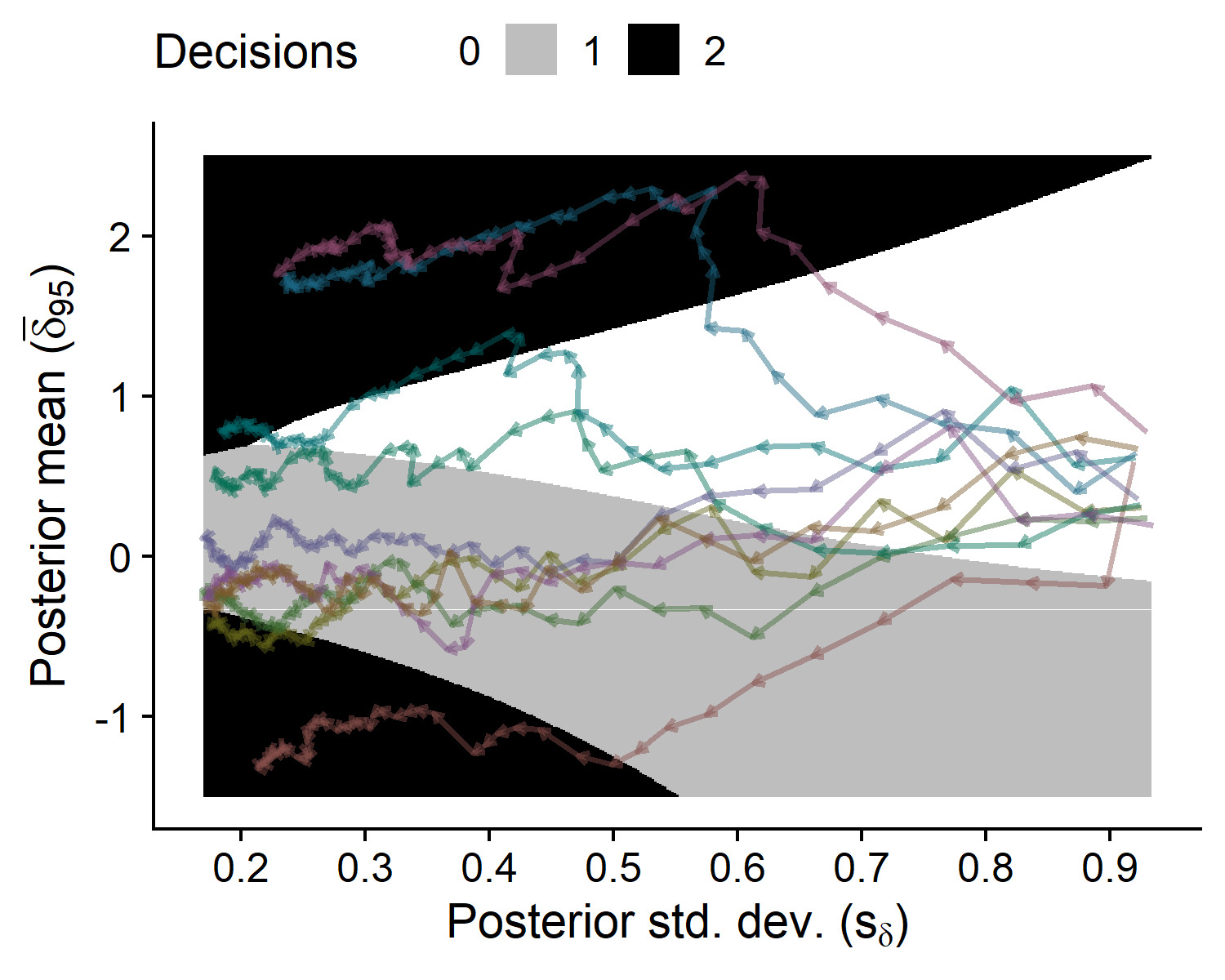}
  \vspace*{-0.8cm}
  \caption{RL: policy gradients}
  \label{fig:Eg2_bound:rl}
  \end{subfigure}%
\caption{Optimal decisions in Example 2. Comparison of fitted decision boundaries.} 
\label{fig:Eg2_bound}
\end{figure}

The results are in Figure \ref{fig:Eg2_bound:bsd}.
Again, the sequential decision problem is reduced to the optimization
problem of finding the optimal $\phi$ in \eqref{eq:Qhphi}.
Since now $\phi \in \Re^3$ the evaluation of $\Qh$ requires a
3-dimensional grid.
 To borrow strength from Monte Carlo evaluations of
\eqref{eq:Qhphi} across neighboring grid points for $\phi$ we proceed
as follows.
We evaluate $\Qh(\phi)$ on a coarse $10 \times 10 \times 10$ grid, and
then fit a quadratic response surface (as a function of $\phi$) to
these Monte Carlo estimates. 
The optimal decision $\phi^*$ is the maximizer of the quadratic
fit. We find $\phi^* = (b_1^*, b_2^*, c^*) = (1.572, 1.200, 0.515)$
 Instead of evaluating $\Qh$ on a regular grid over $\phi$, one
could alternatively select a random number of design points (in
$\phi$). 

 The use of 
parametric boundaries is closely related to
the notion of function approximation and the method of policy gradients in RL, which will be described next.

\section{Reinforcement learning}\label{sec:rl}

The basic setup in RL is usually framed in terms of Markov decision process
(MDP) \citep{puterman2014markov}. The Markov property ensures that 
optimal decisions depend only on the most recently observed state,
enabling practicable algorithms.
In this section we first describe MDPs and
how a sequential design problem can be adapted to fit in this
framework.
Next, we discuss two algorithms, Q-learning \citep{watkins1992q} and
policy gradients 
\citep{grondman2012survey}, implemented in Example 1 and 2,
respectively. Both methods are implemented using neural
networks. Throughout this section, the summary statistics $S_t$ are
referred to as \emph{states},  in keeping with  the common
terminology in the RL literature.

\begin{figure}[hbtp]
  \begin{center}
    \centering
    \begin{subfigure}[t]{.48\linewidth}\centering
    \includegraphics[width=\textwidth]{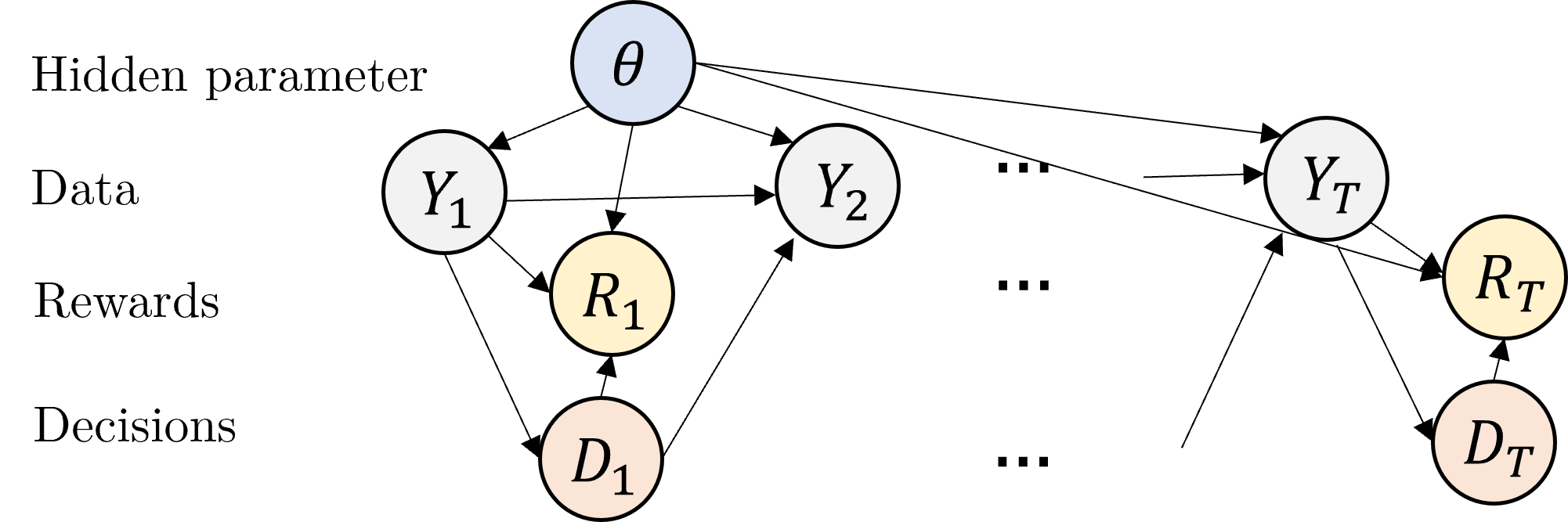}
    \caption{SDP as generic HiMDP.}  
    \label{fig:diagrams:mdp}
    \end{subfigure}\hfill%
    \begin{subfigure}[t]{.48\linewidth}\centering
    \includegraphics[width=\textwidth]{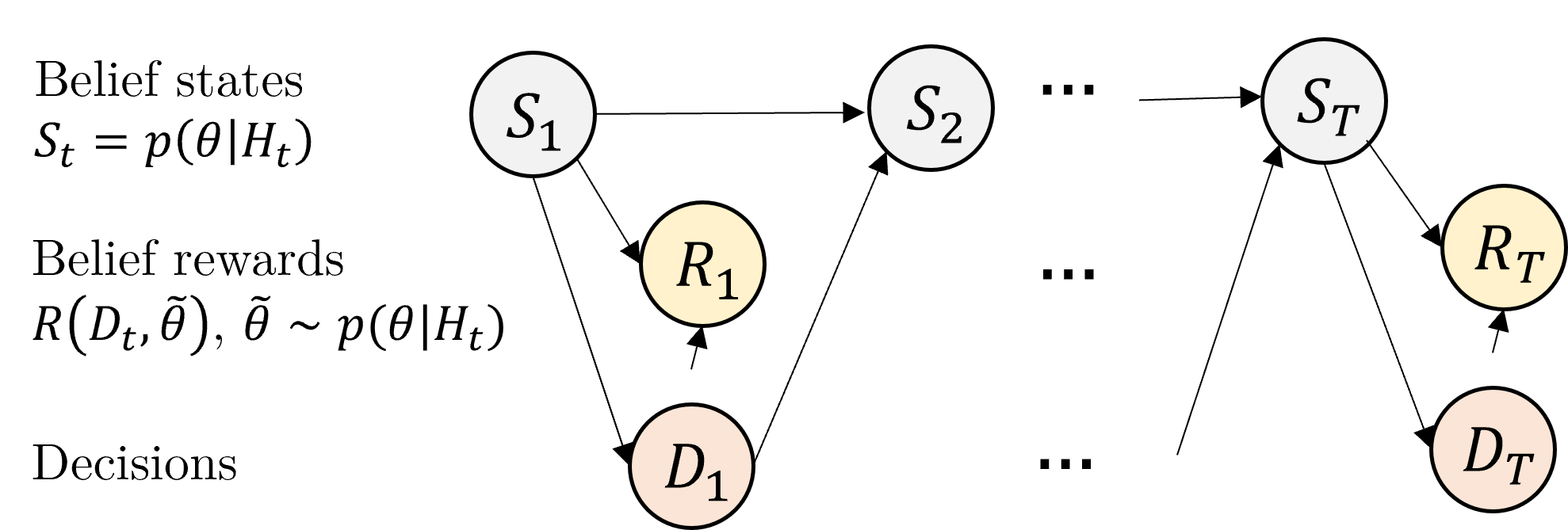}
    \caption{Resulting belief MDP.} 
    \label{fig:diagrams:rl} 
    \end{subfigure}
    \caption{Belief MDP for the SDP.}
    \label{fig:diagrams:mdps}
  \end{center}
\end{figure}

\subsection{Markov decision processes and partial observability}

The Markov property for a decision process  is defined by the
conditions 
$$
p(S_{t+1}\mid H_t, D_t)=p(S_{t+1}\mid S_t, D_t)\quad\quad \mbox{ and }
p(R_{t} \mid H_{t}, D_t)=p(R_{t} \mid S_t, D_t),
$$ That is, the next state $S_{t+1}$ and the reward $R_t$
depend on the history only indirectly through the
current state and action.  When the condition
holds, the decision process is called an MDP. For MDPs, the optimal
policy is only a function of the latest state $S_t$ and not of the
entire history $H_t$ \citep{puterman2014markov}.
Many 
RL algorithms assume the Markov property.
However, many sequential decision problems are more naturally
characterized as
\textit{partially observable MDP} (POMDP) that satisfy the Markov
property only conditional on some $\th$ that is generated at the
beginning of each episode.
Such problems  have

been studied in the RL literature under the name of Hidden-Parameter
MDP (HiMDP) \citep{doshi2016hidden}.

There is a standard -- Bayesian motivated -- way to cast any POMDP as
an MDP using so-called \emph{belief states}
\citep{cassandra1997incremental}. Belief states are obtained by
including the posterior  distribution of unobserved parameters
 as a part of the state.
With a slight abuse of notation, we may write the belief
states as $S_t = p(\theta \mid H_t)$. While the belief state is, in general,
a function, it can often be represented as a vector when the
posterior admits a finite (sufficient) summary statistic. The
 reward distribution can also be written in terms of such belief
states as 
\begin{equation}
p(R\mid S_t, D_t) = \int_\theta p(R \mid D_t, \theta)\, dp(\theta \mid
H_t).
\label{eq:RSD}
\end{equation}
See Figure 
\ref{fig:diagrams:mdps} for a graphical representation of an
HiMDP and the resulting belief MDP. 
 
We implement this approach for Example 1. The reward is
chosen to match the definition of utility. It suffices to define it in
terms of $\theta$ (and let the posterior take care of the rest, using
\eqref{eq:RSD}). 
We use 
\begin{equation}\label{eq:reward1}
  R(d, \theta) = -c\,\mathbb{I}(d=0)  - K \mathbb{I}(\theta_d \neq \theta, d\neq 0),
\end{equation}
as in \eqref{eq:utility1}.
Next we introduce the belief states.
Recall the notation from Example 1. We have $p(\th \mid S_t) = \Bin(p_t,
p_t(1-p_t)/t)$. 
The summary statistic $S_t=(t, p_t)$ is a two-dimensional
representation of the belief state.  

 Considering Example 2, we note that
the utility function \eqref{eq:utility2} depends on the state $S_t$
only, and does not involve $\theta$.
We define 
\begin{equation}\label{eq:reward2}
R_\theta(d, S_t) = -c_1 \mathbb{I}(d=0) + (-c_2  \N(S_t) + K  \Delta_R(S_t))  \mathbb{I}(d=2). \\
\end{equation}
While the reward is clearly Markovian, the transition probability is
not necessarily Markovian.  This is the case because in this example
the posterior moments $S_t$ are not a sufficient statistic. 
In practice, however,
 a minor violation of the Markov assumption
for the transition distribution does not seem to affect the
ability to obtain good policies with standard RL techniques.

\subsection{Q-learning} 

Q-learning \citep{watkins1992q, CliftonLaber:20, murphy2003optimal} is
an RL algorithm that is similar in spirit to the constrained backward
induction described in Section \ref{sec:CBI}. The starting
point is Bellman optimality equation for MDPs
\citep{bellman1966dynamic}.  Equation \eqref{eq:bellman-bsd} for
the optimal $\Ds$ and written for MDPs becomes 
\begin{equation}\label{eq:bellman-rl}
  Q^\pis(s, d)=\E\{R_t + \textstyle{\max_{d'}} Q^\pis(S_{t+1}, d')\mid S_t=s, D_t=d\},
\end{equation}
 where the expectation is with respect to $R_t$ and $S_{t+1}$.
The optimal policy is implicitly defined as the solution to
\eqref{eq:bellman-rl}. 

Q-learning proceeds iteratively following the fixed-point iteration
principle. Let $Q^{(k)}$ be some approximation of
$Q^\star$.
We assume that a set of simulated transitions
$\{(s_t,d_t,r_{t},s_{t+1})\}_{t=1}^n$ is available.
This collection is used like the forward simulations in the earlier
discussion. In RL it is known as the ``experience replay buffer'', and can be generated using 
any 
 stochastic policy. And suppose,
 for the moment,  that the state and action spaces are
finite discrete,  allowing to record $Q^{(k)}$ in a table. 
 Q-learning is defined by updating $Q^{(k)}$ as 
\begin{equation}\label{eq:qlearn}
Q^{(k + 1)}(s_t, d_t) \leftarrow (1-\alpha_k)Q^{(k)}(s_t, d_t) +
\alpha_k \{r_t + \textstyle{\max_d} Q^{(k)}(s_{t+1},d)\}. 
\end{equation}
 Note the moving average nature of the update.
The procedure iterates until convergence from a stream of
transitions.

Deep Q-networks (DQN) \citep{mnih2013playing} are an extension of
Q-learning for continuous states.
 A neural network is used to represent $Q(\cdot)$. Let $\phi^{(k)}$
denote the parameters of the neural network at iteration $k$. 
 Using simulated transitions from the buffer 

DQN performs updates
\begin{equation}\label{eq:dqn}
\phi^{(k+1)} = \textstyle{\argmin_\phi \sum_{t=1}^n} \left(r_t + \max_d Q_{\phi^{(k)}}(S_{t+1}, d) - Q_\phi(S_t, d_t)\right)^2.
\end{equation}
In practice, exact minimization is replaced by a gradient step from mini-batches, together with numerous implementation tricks \citep{mnih2013playing}.

We implemented DQN  in Example 1 using the Python package {\tt
  Stable-Baselines3} \citep{stable-baselines3}.
 The experience replay buffer is continuously updated.
The algorithm uses a random policy to produce an initial buffer and then
adds experience from an $\epsilon$-greedy policy, where the current
best guess for the optimal policy is chosen with probability
$(1-\epsilon)$, and otherwise fully random actions are chosen with
probability $\epsilon$. 
Figure
\ref{fig:Eg1-value:rl} shows $\hat{Q}$, the estimate of $\Qs$, for
each state and action $d\in\{0,1,2\}$. Figure
\ref{fig:Eg1-value:rl-bounds} shows the corresponding optimal
actions. Overall, the results are similar to
 the results with constrained backward induction, but much
smoother. 
Also, notice that the solution under DQN is usually better
in terms of expected utility
 as shown in Figure \ref{fig:Eg1:parametric-values},
even in (out-of-sample) evaluation episodes.
This is likely due to to the flexibility and high-dimensional nature of
the neural network approximation.

The better performance of RL comes at a price. First is
sample efficiency (the number of simulations required by the algorithm to yield a good policy). The best
$\hat{Q}$ is obtained after 2 million sampled transitions.
Data efficiency is a known problem in
DQN, and in RL in general \citep{yu2018towards}. In many real
applications investigators cannot afford such a high number of simulation
steps. Another limitation is training instability. In particular, Figure
\ref{fig:Eg1:parametric-values} illustrates a phenomenon known as
\emph{catastrophic forgetting}, which happens when additional training
decreases the ability of the agent to perform a previously learned
task \citep{atkinson2021pseudo}. 
 This can happen because of the instability that arises
from a typical strategy of evaluating performance periodically and
keeping track of the best performing policy only. 
Several improvements over basic DQN have been proposed,
with improved performance and efficiency
\citep{hessel2018rainbow}.

\subsection{Policy gradients} 
The approach is similar to the use of
parametric boundaries discussed before.
Policy gradient (PG) approaches start from
a parameterization $\pi_\vphi$ of the policy. Again, consider a neural network (NN)  with weights  $\phi$. The goal of a PG
method is to maximize the objective 
\begin{equation}\label{eq:pg-objective}
\max_\phi J(\phi)=\E\{G \mid \pi_\phi\}.
\end{equation}
This objective is the analogue to maximizing $\UQ(\phi)$ in
\eqref{eq:Qhphi}, except that here the stochastic policy
$\pi_\phi$ is a probability distribution on decisions $S_t$.
The main characteristic of PG methods is the use of 
gradient descent to solve \eqref{eq:pg-objective}.

The evaluation of gradients is based on the PG theorem
\citep{sutton1999policy}, 
\begin{equation}\label{eq:policy-grad}
  \nabla_\vphi J(\vphi)=\E\left\{
    \left[\sum_{t=1}^T \nabla_\vphi \log \pi_\vphi(D_t \mid
      S_t)\right] \cdot
    G \mid \pi_\phi \right\},
\end{equation}
where the total return $G$ is a function $G(\tau)$ of the entire
trajectory $\tau=(S_1, D_1,\hdots, S_T, D_T)$.
Using gradients, PG methods can optimize over high-dimensional parameter
spaces like in neural networks. In practice, estimates of the gradient
are known to have huge variance, affecting the optimization. But
several implementation tricks exist that improve the stability and
reduce the variance. Proximal policy optimization (PPO)
\citep{schulman2017proximal} incorporates many of these tricks, and is
widely used as a default for PG-based methods. 

 The PG theorem is essentially Leibniz rule for the gradient
of $J(\phi)$.

With a slight abuse of notation, write $\pi_\vphi(\tau)$ for
the distribution of $\tau$ induced by $\pi_\vphi$  for the
sequential decisions. 
Then Leibniz rule for the gradient of the integral gives
\begin{multline}
  \nabla_\vphi J(\vphi) =
 \nabla_\vphi\int \pi_\phi(\tau) G(\tau) d\tau = \\
\int  \nabla_\vphi \pi_\vphi(\tau) \cdot G(\tau)\, d\tau
=
\int \left(\nabla_\vphi \log \pi_\vphi(\tau)\right)\,
\pi_\vphi(\tau) \cdot G(\tau)\, d\tau =
\E \left\{ \nabla_\vphi \log \pi_\vphi(\tau) \cdot G(\tau)\right\},
\nonumber
\end{multline}
where the expectations are with respect to the (stochastic) policy
$\pi_\phi$ over $\tau$. 
The log probability in the last expression can be written as a sum of
log probabilities, yielding \eqref{eq:policy-grad}. 

 PPO is implemented in Example 2 using {\tt Stable-Baselines3}
\citep{stable-baselines3}.  The results are shown in Figure
\ref{fig:Eg2_bound:rl}. Not surprisingly, the results are similar
 to those obtained earlier using parametric decision boundaries. 
Interestingly, the figure shows that neural networks do not
necessarily extrapolate 
well to regions with low data. This behavior is noticeable on the
lower left corner of the figure, where there \emph{could} be data, but
where it is never observed in practice because of the early stopping
implied by the boundaries.

\section{Discussion}
We have introduced some of the main features of RL and BSD in the
context of two optimal stopping problems.
In the context of these examples the two approaches are quite similar,
including an almost one-to-one mapping of terminology and notation, as
we attempted in Table \ref{tab:notation}.
In general, however, the applicability, especially the practical use
of RL is much wider. The restriction of the sequential problems to
optimal stopping was only needed for easy application of the BSD
solution. In contrast, RL methods are routinely used for a variety of
other problems, such as robotics \citep{tunyasuvunakool2020dm_control} autonomous driving \citep{sallab2017deep,wurman2022outracing}, and smart building energy management \citep{yu2021review}. The main attraction of BSD is the principled nature of the
solution. One can argue from first principles that a rational agent
should act as if he or she were optimizing expected utility as in
\eqref{eq:DTs}. There is a well-defined and coherent propagation of
uncertainties. This might be particularly important when the
SDP and underlying model are only part of a bigger problem.
Overall, we note that the perspective of one method, and
corresponding algorithms can be useful for improvements in the
respective other method. For example, policy gradients could readily
be used to solve BSD if randomized decision rules were used. The
latter is usually not considered.
On the other hand, hierarchical Bayesian inference models could be
used to combine multiple sources of evidence in making sequential
decisions under RL, or multiple related problems could be linked in a
well-defined manner in a larger encompassing model. For example,
clinical trials are never carried out in isolation. Often the same
department or group might run multiple trials on the same patient
population for the same disease, with obvious opportunities to borrow
strength. 

\section*{Acknowledgements}
\thanks{
The authors gratefully thank Peter Stone and the Learning Agents Research Group (LARG) for helpful discussions and feedback.}

\section*{Funding}
Yunshan Duan and Peter M\"uller are partially funded by the NSF under grant NSF/DMS 1952679. 

\section*{Conflicts of Interest}
The authors report there are no competing interests to declare

\clearpage \newpage

\bibliographystyle{Chicago}
\bibliography{bibfile}

\clearpage \newpage
\appendix

\section*{Appendix} 

\section{Details in Example 2}
We assume a nonlinear regression sampling model
$$Y_t = f(X_t \mid \theta) + \epsilon_t, \;\;\;\; \epsilon_t \sim N(0,\sigma^2),$$
and the dose-response curve
\begin{equation*}
  f(X_t \mid \theta)  = a + b \frac{X_t^r}{(q^r + X_t^r)}
\end{equation*}
We fix $a,r,\sigma^2$, and put a normal prior on unknown parameters $\theta = (b,q)$,
$$p(\theta) = N(\theta_0,diag(\lambda_0)).$$
\paragraph{Sample size calculation}
If at time $T$, the decision $D_T=2$ indicates stopping and a pivotal trial is conducted to test $H_0: \delta_{95}=0$ vs. $H_1: \delta_{95}>0$. We need to determine the sample size $N_{\alpha,\beta}$ for the pivotal trial that can achieve desired significance level $\alpha$ and power $(1-\beta)$, and calculate the posterior predictive probability of a significant outcome, $\Delta_R = \Pr(\mbox{reject $H_0$ in the 2nd trial} \mid H_T)$.

Let $\dhat$ and $s_{\delta}$ denote the posterior mean and std. dev. of $\delta_{95}$. We calculate power based on $\delta_{95} = \delta^*$, where $\delta_{95}^* = \dhat - s_{\delta}$.

Now consider a test enrolling $N_{\alpha,\beta}$ patients, randomizing $N_{\alpha,\beta}/2$ at $x=0$
  (placebo) and $N_{\alpha,\beta}/2$ at the estimated ED95.
  Assuming $var(y_i)=1$, we need
  $$
  N_{\alpha,\beta} \ge 4 \left[ (q_\alpha+ q_\beta)/\delta^* \right]^2
  $$
  where $q_\alpha$ is the $\alpha$ right tail cutoff for the $N(0,1)$
  and $\alpha=5\%$ and $(1-\beta)=80\%$ are the desired significance level
  and power (i.e., $\beta=0.2$).
  
  A significant outcome at the end of the 2nd trial means
  data in the rejection region. Let $\ybar_1, \ybar_0$ denote the sample
  average of $N_{\alpha,\beta}/2$ patients each to be enrolled in the two arms of
  the 2nd trial. Then the rejection region is
  $$
  R = \{(\ybar_1-\ybar_0)\sqrt{N_{\alpha,\beta}/4} \ge q_\alpha\}.
  $$
  Let $\Phi(\cdot)$ denote the standard normal c.d.f. Then
  $$
  \Delta_R = \Phi\left[
    \frac{\bar{\delta} \sqrt{N_{\alpha,\beta}/4} - q_\alpha}
    {\sqrt{1+\frac{N_{\alpha,\beta}}4 s_\delta^2}} \right]
  $$

\paragraph{Posterior simulation}
We can implement independent posterior simulation:
\begin{enumerate}[(i)]
    \item Generate $q \sim p(q \mid H_t)$, using
    $$
p(q \mid H_t) \propto
p(H_t \mid q) \cdot
p(q)
= \frac{p(H_t \mid b, q) p(b)}{p(b \mid q,H_t)} \cdot p(q)
$$
    \item Then generate $b$ from the posterior conditional distribution $b \sim p(b \mid q, H_t)$. Based on normal linear regression, the conditional posterior is a univariate normal distribution.
\end{enumerate}

\end{document}

%% file: math.tex
\usepackage{amsmath,amsfonts,bm}

\def\eqref#1{equation~\ref{#1}}

\def\1{\bm{1}}

\def\rx{{\textnormal{x}}}

\def\vphi{{\bm{\phi}}}

\DeclareMathAlphabet{\mathsfit}{\encodingdefault}{\sfdefault}{m}{sl}
\SetMathAlphabet{\mathsfit}{bold}{\encodingdefault}{\sfdefault}{bx}{n}

\def\gP{{\mathcal{P}}}

\def\sX{{\mathbb{X}}}

\newcommand{\E}{\mathbb{E}}

\DeclareMathOperator*{\argmax}{arg\,max}
\DeclareMathOperator*{\argmin}{arg\,min}

%% file: main.bbl
\begin{thebibliography}{}

\bibitem[\protect\citeauthoryear{Atkinson, McCane, Szymanski, and
  Robins}{Atkinson et~al.}{2021}]{atkinson2021pseudo}
Atkinson, C., B.~McCane, L.~Szymanski, and A.~Robins (2021).
\newblock Pseudo-rehearsal: Achieving deep reinforcement learning without
  catastrophic forgetting.
\newblock {\em Neurocomputing\/}~{\em 428}, 291--307.

\bibitem[\protect\citeauthoryear{Bellman}{Bellman}{1966}]{bellman1966dynamic}
Bellman, R. (1966).
\newblock Dynamic programming.
\newblock {\em Science\/}~{\em 153\/}(3731), 34--37.

\bibitem[\protect\citeauthoryear{Berger}{Berger}{2013}]{berger2013statistical}
Berger, J.~O. (2013).
\newblock {\em Statistical decision theory and {B}ayesian analysis}.
\newblock Springer Science \& Business Media.

\bibitem[\protect\citeauthoryear{Berry and Ho}{Berry and Ho}{1988}]{BerryHo88}
Berry, D.~A. and C.-H. Ho (1988).
\newblock {One-sided sequential stopping boundaries for clinical trials: A
  decision-theoretic approach}.
\newblock {\em Biometrics\/}~{\em 44\/}(1), 219--227.

\bibitem[\protect\citeauthoryear{Brockwell and Kadane}{Brockwell and
  Kadane}{2003}]{BrockwellKadane03}
Brockwell, A.~E. and J.~B. Kadane (2003).
\newblock A gridding method for {B}ayesian sequential decision problems.
\newblock {\em Journal of Computational and Graphical Statistics\/}~{\em
  12\/}(3), 566--584.

\bibitem[\protect\citeauthoryear{Carlin, Kadane, and Gelfand}{Carlin
  et~al.}{1998}]{carlin1998approaches}
Carlin, B.~P., J.~B. Kadane, and A.~E. Gelfand (1998).
\newblock Approaches for optimal sequential decision analysis in clinical
  trials.
\newblock {\em Biometrics\/}~{\em 54}, 964--975.

\bibitem[\protect\citeauthoryear{Cassandra, Littman, and Zhang}{Cassandra
  et~al.}{1997}]{cassandra1997incremental}
Cassandra, A., M.~L. Littman, and N.~L. Zhang (1997).
\newblock {Incremental pruning: {A} simple, fast, exact method for partially
  observable {M}arkov decision processes}.
\newblock In {\em Proceedings of the Thirteenth conference on Uncertainty in
  artificial intelligence}, pp.\  54--61.

\bibitem[\protect\citeauthoryear{Christen and Nakamura}{Christen and
  Nakamura}{2003}]{ChristenNakamura03}
Christen, J.~A. and M.~Nakamura (2003).
\newblock Sequential stopping rules for species accumulation.
\newblock {\em Journal of Agricultural, Biological, and Environmental
  Statistics\/}~{\em 8\/}(2), 184--195.

\bibitem[\protect\citeauthoryear{Clifton and Laber}{Clifton and
  Laber}{2020}]{CliftonLaber:20}
Clifton, J. and E.~Laber (2020).
\newblock {Q-learning: Theory and applications}.
\newblock {\em Annual Review of Statistics and Its Application\/}~{\em 7\/}(1),
  279--301.

\bibitem[\protect\citeauthoryear{DeGroot}{DeGroot}{2004}]{degr:70}
DeGroot, M. (2004).
\newblock {\em Optimal statistical decisions}.
\newblock New York: Wiley-Interscience.

\bibitem[\protect\citeauthoryear{Doshi-Velez and Konidaris}{Doshi-Velez and
  Konidaris}{2016}]{doshi2016hidden}
Doshi-Velez, F. and G.~Konidaris (2016).
\newblock Hidden parameter {Markov} decision processes: A semiparametric
  regression approach for discovering latent task parametrizations.
\newblock In {\em IJCAI: Proceedings of the Conference}, Volume 2016, pp.\
  1432.

\bibitem[\protect\citeauthoryear{Gaon and Brafman}{Gaon and
  Brafman}{2020}]{gaon2020reinforcement}
Gaon, M. and R.~Brafman (2020).
\newblock Reinforcement learning with non-{Markovian} rewards.
\newblock In {\em Thirty-fourth AAAI Conference on Artificial Intelligence}.

\bibitem[\protect\citeauthoryear{Grieve and Krams}{Grieve and
  Krams}{2005}]{grieve2005astin}
Grieve, A.~P. and M.~Krams (2005).
\newblock {ASTIN}: {A} {B}ayesian adaptive dose--response trial in acute
  stroke.
\newblock {\em Clinical Trials\/}~{\em 2\/}(4), 340--351.

\bibitem[\protect\citeauthoryear{Grondman, Busoniu, Lopes, and
  Babuska}{Grondman et~al.}{2012}]{grondman2012survey}
Grondman, I., L.~Busoniu, G.~A. Lopes, and R.~Babuska (2012).
\newblock A survey of actor-critic reinforcement learning: Standard and natural
  policy gradients.
\newblock {\em IEEE Transactions on Systems, Man, and Cybernetics, Part C
  (Applications and Reviews)\/}~{\em 42\/}(6), 1291--1307.

\bibitem[\protect\citeauthoryear{Hessel, Modayil, Van~Hasselt, Schaul,
  Ostrovski, Dabney, Horgan, Piot, Azar, and Silver}{Hessel
  et~al.}{2018}]{hessel2018rainbow}
Hessel, M., J.~Modayil, H.~Van~Hasselt, T.~Schaul, G.~Ostrovski, W.~Dabney,
  D.~Horgan, B.~Piot, M.~Azar, and D.~Silver (2018).
\newblock Rainbow: Combining improvements in deep reinforcement learning.
\newblock In {\em Thirty-second AAAI Conference on Artificial Intelligence}.

\bibitem[\protect\citeauthoryear{Kadane and Vlachos}{Kadane and
  Vlachos}{2002}]{KadaneVlachos02}
Kadane, J.~B. and P.~K. Vlachos (2002).
\newblock Hybrid methods for calculating optimal few-stage sequential
  strategies: Data monitoring for a clinical trial.
\newblock {\em Statistics and Computing\/}~{\em 12}, 147--152.

\bibitem[\protect\citeauthoryear{MacDonald, Ranjan, and Chipman}{MacDonald
  et~al.}{2015}]{macdonald2015gpfit}
MacDonald, B., P.~Ranjan, and H.~Chipman (2015).
\newblock {GPfit}: An {R} package for fitting a {G}aussian process model to
  deterministic simulator outputs.
\newblock {\em Journal of Statistical Software\/}~{\em 64}, 1--23.

\bibitem[\protect\citeauthoryear{Meibohm and Derendorf}{Meibohm and
  Derendorf}{1997}]{meibohm1997basic}
Meibohm, B. and H.~Derendorf (1997).
\newblock Basic concepts of pharmacokinetic/pharmacodynamic (pk/pd) modelling.
\newblock {\em International Journal of Clinical Pharmacology and
  Therapeutics\/}~{\em 35\/}(10), 401--413.

\bibitem[\protect\citeauthoryear{Mnih, Kavukcuoglu, Silver, Graves, Antonoglou,
  Wierstra, and Riedmiller}{Mnih et~al.}{2013}]{mnih2013playing}
Mnih, V., K.~Kavukcuoglu, D.~Silver, A.~Graves, I.~Antonoglou, D.~Wierstra, and
  M.~Riedmiller (2013).
\newblock {Playing Atari with deep reinforcement learning}.
\newblock arXiv preprint 1312.5602.

\bibitem[\protect\citeauthoryear{M{\"u}ller, Berry, Grieve, Smith, and
  Krams}{M{\"u}ller et~al.}{2007}]{muller2007simulation}
M{\"u}ller, P., D.~A. Berry, A.~P. Grieve, M.~Smith, and M.~Krams (2007).
\newblock Simulation-based sequential {B}ayesian design.
\newblock {\em Journal of Statistical Planning and Inference\/}~{\em
  137\/}(10), 3140--3150.

\bibitem[\protect\citeauthoryear{Murphy}{Murphy}{2003}]{murphy2003optimal}
Murphy, S.~A. (2003).
\newblock Optimal dynamic treatment regimes.
\newblock {\em Journal of the Royal Statistical Society: Series B (Statistical
  Methodology)\/}~{\em 65\/}(2), 331--355.

\bibitem[\protect\citeauthoryear{Murphy, Oslin, Rush, and Zhu}{Murphy
  et~al.}{2007}]{murphy2007methodological}
Murphy, S.~A., D.~W. Oslin, A.~J. Rush, and J.~Zhu (2007).
\newblock Methodological challenges in constructing effective treatment
  sequences for chronic psychiatric disorders.
\newblock {\em Neuropsychopharmacology\/}~{\em 32\/}(2), 257--262.

\bibitem[\protect\citeauthoryear{Parmigiani and Inoue}{Parmigiani and
  Inoue}{2009}]{ParmigianiLurdes:09}
Parmigiani, G. and L.~Inoue (2009).
\newblock {\em Decision Theory: Principles and Approaches}.
\newblock Wiley.

\bibitem[\protect\citeauthoryear{Puterman}{Puterman}{2014}]{puterman2014markov}
Puterman, M.~L. (2014).
\newblock {\em {Markov decision processes: {D}iscrete stochastic dynamic
  programming}}.
\newblock John Wiley \& Sons.

\bibitem[\protect\citeauthoryear{Raffin, Hill, Gleave, Kanervisto, Ernestus,
  and Dormann}{Raffin et~al.}{2021}]{stable-baselines3}
Raffin, A., A.~Hill, A.~Gleave, A.~Kanervisto, M.~Ernestus, and N.~Dormann
  (2021).
\newblock Stable-baselines3: Reliable reinforcement learning implementations.
\newblock {\em Journal of Machine Learning Research\/}~{\em 22\/}(268), 1--8.

\bibitem[\protect\citeauthoryear{Robins}{Robins}{1997}]{robins1997causal}
Robins, J.~M. (1997).
\newblock Causal inference from complex longitudinal data.
\newblock In {\em Latent variable modeling and applications to causality}, pp.\
   69--117. Springer.

\bibitem[\protect\citeauthoryear{Rossell, M\"uller, and Rosner}{Rossell
  et~al.}{2007}]{Rossell:13b}
Rossell, D., P.~M\"uller, and G.~Rosner (2007).
\newblock Screening designs for drug development.
\newblock {\em Biostatistics\/}~{\em 8}, 595--608.

\bibitem[\protect\citeauthoryear{Sallab, Abdou, Perot, and Yogamani}{Sallab
  et~al.}{2017}]{sallab2017deep}
Sallab, A.~E., M.~Abdou, E.~Perot, and S.~Yogamani (2017).
\newblock Deep reinforcement learning framework for autonomous driving.
\newblock {\em Electronic Imaging\/}~{\em 2017\/}(19), 70--76.

\bibitem[\protect\citeauthoryear{Schulman, Wolski, Dhariwal, Radford, and
  Klimov}{Schulman et~al.}{2017}]{schulman2017proximal}
Schulman, J., F.~Wolski, P.~Dhariwal, A.~Radford, and O.~Klimov (2017).
\newblock Proximal policy optimization algorithms.
\newblock arXiv preprint 1707.06347.

\bibitem[\protect\citeauthoryear{Shen and Huan}{Shen and
  Huan}{2021}]{shen2021bayesian}
Shen, W. and X.~Huan (2021).
\newblock Bayesian sequential optimal experimental design for nonlinear models
  using policy gradient reinforcement learning.
\newblock arXiv preprint 2110.15335.

\bibitem[\protect\citeauthoryear{Silver, Hubert, Schrittwieser, Antonoglou,
  Lai, Guez, Lanctot, Sifre, Kumaran, Graepel, et~al.}{Silver
  et~al.}{2018}]{silver2018general}
Silver, D., T.~Hubert, J.~Schrittwieser, I.~Antonoglou, M.~Lai, A.~Guez,
  M.~Lanctot, L.~Sifre, D.~Kumaran, T.~Graepel, et~al. (2018).
\newblock A general reinforcement learning algorithm that masters chess, shogi,
  and go through self-play.
\newblock {\em Science\/}~{\em 362\/}(6419), 1140--1144.

\bibitem[\protect\citeauthoryear{Silver, Lever, Heess, Degris, Wierstra, and
  Riedmiller}{Silver et~al.}{2014}]{silver2014deterministic}
Silver, D., G.~Lever, N.~Heess, T.~Degris, D.~Wierstra, and M.~Riedmiller
  (2014).
\newblock Deterministic policy gradient algorithms.
\newblock In {\em Proceedings of the 31st International Conference on
  International Conference on Machine Learning}, pp.\  387--395.

\bibitem[\protect\citeauthoryear{Sutton and Barto}{Sutton and
  Barto}{2018}]{sutton2018reinforcement}
Sutton, R.~S. and A.~G. Barto (2018).
\newblock Reinforcement learning: An introduction.
\newblock {\em MIT Press\/}~{\em 5}, 31.

\bibitem[\protect\citeauthoryear{Sutton, McAllester, Singh, and Mansour}{Sutton
  et~al.}{1999}]{sutton1999policy}
Sutton, R.~S., D.~McAllester, S.~Singh, and Y.~Mansour (1999).
\newblock Policy gradient methods for reinforcement learning with function
  approximation.
\newblock In S.~Solla, T.~Leen, and K.~M\"{u}ller (Eds.), {\em Advances in
  Neural Information Processing Systems}, Volume~12.

\bibitem[\protect\citeauthoryear{Tunyasuvunakool, Muldal, Doron, Liu, Bohez,
  Merel, Erez, Lillicrap, Heess, and Tassa}{Tunyasuvunakool
  et~al.}{2020}]{tunyasuvunakool2020dm_control}
Tunyasuvunakool, S., A.~Muldal, Y.~Doron, S.~Liu, S.~Bohez, J.~Merel, T.~Erez,
  T.~Lillicrap, N.~Heess, and Y.~Tassa (2020).
\newblock dm\_control: Software and tasks for continuous control.
\newblock {\em Software Impacts\/}~{\em 6}, 100022.

\bibitem[\protect\citeauthoryear{Watkins and Dayan}{Watkins and
  Dayan}{1992}]{watkins1992q}
Watkins, C.~J. and P.~Dayan (1992).
\newblock Q-learning.
\newblock {\em Machine learning\/}~{\em 8\/}(3-4), 279--292.

\bibitem[\protect\citeauthoryear{Wurman, Barrett, Kawamoto, MacGlashan,
  Subramanian, Walsh, Capobianco, Devlic, Eckert, Fuchs, et~al.}{Wurman
  et~al.}{2022}]{wurman2022outracing}
Wurman, P.~R., S.~Barrett, K.~Kawamoto, J.~MacGlashan, K.~Subramanian, T.~J.
  Walsh, R.~Capobianco, A.~Devlic, F.~Eckert, F.~Fuchs, et~al. (2022).
\newblock Outracing champion gran turismo drivers with deep reinforcement
  learning.
\newblock {\em Nature\/}~{\em 602\/}(7896), 223--228.

\bibitem[\protect\citeauthoryear{Yu, Qin, Zhang, Shen, Jiang, and Guan}{Yu
  et~al.}{2021}]{yu2021review}
Yu, L., S.~Qin, M.~Zhang, C.~Shen, T.~Jiang, and X.~Guan (2021).
\newblock A review of deep reinforcement learning for smart building energy
  management.
\newblock {\em IEEE Internet of Things Journal\/}~{\em 8}, 12046--12063.

\bibitem[\protect\citeauthoryear{Yu}{Yu}{2018}]{yu2018towards}
Yu, Y. (2018).
\newblock Towards sample efficient reinforcement learning.
\newblock In {\em Proceedings of the Twenty-Seventh International Joint
  Conference on Artificial Intelligence}.

\bibitem[\protect\citeauthoryear{Zhao, Kosorok, and Zeng}{Zhao
  et~al.}{2009}]{zhao2009reinforcement}
Zhao, Y., M.~R. Kosorok, and D.~Zeng (2009).
\newblock Reinforcement learning design for cancer clinical trials.
\newblock {\em Statistics in Medicine\/}~{\em 28\/}(26), 3294--3315.

\end{thebibliography}
